\documentclass[floats,floatfix,amssymb,prd,twocolumn,superscriptaddress,nofootinbib]{revtex4-1}

\usepackage{subcaption}
\usepackage{ragged2e}
\DeclareCaptionJustification{justified}{\justifying}
\makeatletter
\newcommand{\subsetsim}{\mathrel{\mathpalette\subset@sim\relax}}
\newcommand{\subset@sim}[2]{%
  \vtop{\offinterlineskip\m@th
    \ialign{\hfil##\cr
      $#1\subset$\cr\noalign{\kern0.5pt}\scalebox{0.9}{$#1\sim$}\cr
    }%
  }%
}
\makeatother

\usepackage{amssymb,amsmath,verbatim,mathtools,needspace,enumitem,etoolbox,graphicx,physics,microtype,afterpage,bm}
\usepackage[dvipsnames, usenames]{xcolor}
\definecolor{linkcolor}{rgb}{0.0,0.3,0.5}
\usepackage{booktabs}

\definecolor{oucrimsonred}{rgb}{0.6, 0.0, 0.0}
\definecolor{persianblue}{rgb}{0.11, 0.22, 0.73}
\definecolor{forestgreen}{rgb}{0.13,0.35,0.13}
\usepackage[unicode, 
colorlinks=true, 
linkcolor=persianblue, 
citecolor=forestgreen, 
filecolor=persianblue,
urlcolor=oucrimsonred, 
pdfusetitle]{hyperref}

\usepackage[all]{hypcap}
\usepackage[T1]{fontenc}
\usepackage[utf8]{inputenc}
\usepackage{tabularx}
\usepackage{adjustbox}
\usepackage{float}
\usepackage{ulem}
\usepackage{xfrac}
\usepackage{orcidlink}

\definecolor{azure}{rgb}{0.0, 0.5, 1.0}
\definecolor{deepfuchsia}{rgb}{0.76, 0.33, 0.76}
\definecolor{VioletRed4}{rgb}{0.55, 0.13, .32}
 
\definecolor{harvardcrimson}{rgb}{0.79, 0.0, 0.09}
\definecolor{oceanboatblue}{rgb}{0.0, 0.47, 0.75}
\definecolor{persianblue}{rgb}{0.11, 0.22, 0.73}
\definecolor{egyptianblue}{rgb}{0.06, 0.2, 0.65}
\definecolor{navyblue}{rgb}{0.0, 0.0, 0.5}

\usepackage{multirow}
\usepackage{pifont}
\usepackage{fontawesome}
\usepackage{lmodern}

\usepackage{multirow}

\allowdisplaybreaks
\usepackage{tikz}
\usepackage{color}
\usepackage{framed}

\definecolor{rossos}{cmyk}{0,1,1,0.55}
\definecolor{bluscuro}{rgb}{0.15, 0.2, .85}
\definecolor{bluchiaro}{cmyk}{1,.3,0.,0.1}
\definecolor{ForestGreen}{rgb}{0.13, 0.55, 0.13}

\newcommand{\MPl}{\bar{M}_{\textrm{\tiny{Pl}}}}

\def\nn{\nonumber}

\def\bea{\begin{eqnarray}}
\def\eea{\end{eqnarray}}

\newcommand{\bs}{\begin{subequations}}
\newcommand{\es}{\end{subequations}}

\newcommand{\be}{\begin{equation}}
\newcommand{\ee}{\end{equation}}

\def\lsim{\mathrel{\rlap{\lower4pt\hbox{\hskip0.5pt$\sim$}}
    \raise1pt\hbox{$<$}}}         
\def\gsim{\mathrel{\rlap{\lower4pt\hbox{\hskip0.5pt$\sim$}}
    \raise1pt\hbox{$>$}}}         

\makeatletter
\def\l@subsubsection#1#2{}
\makeatother

\newcommand{\sapienza}{Dipartimento di Fisica, Sapienza Università 
	di Roma, Piazzale Aldo Moro 5, 00185, Roma, Italy}
\newcommand{\infn}{INFN, Sezione di Roma, Piazzale Aldo Moro 2, 00185, Roma, Italy}

\begin{document}

\title{
On the inflationary interpretation of the nHz 
gravitational-wave background
}

\author{Lorenzo Frosina}
\email{frosina.1763813@studenti.uniroma1.it}
\affiliation{\sapienza}

\author{Alfredo Urbano\orcidlink{0000-0002-0488-3256}}
\email{alfredo.urbano@uniroma1.it}
\affiliation{\sapienza}
\affiliation{\infn}


\begin{abstract}
We construct a single-field  model of inflation that achieves remarkable agreement with Planck and BICEP/Keck  cosmological observations. The model, via the presence of an ultra-slow-roll phase, generates a sizable scalar-induced gravitational wave (GW) signal at nHz frequencies. We elucidate the distinctive features of this signal concerning its connection to the recent measurement of the low-frequency GW background reported by the NANOGrav collaboration. 
\end{abstract}

\maketitle

{
  \hypersetup{linkcolor=black}
}

\normalem

\section{Introduction}\label{sec:Intro}

In the context of cosmology, inflation is a pivotal concept that explains various fundamental features of the observable universe on large scales. It operates on two levels: background expansion and perturbations. At the level of background expansion, inflation accounts for several critical aspects of the universe's structure, such as its vast size, homogeneity, isotropy, and flatness. On the level of perturbations, inflation plays a significant role in initiating the process of structure formation. It generates, in the form of primordial density fluctuations, the essential ``seeds'' for creating the structure we observe in the universe today, such as galaxies and galaxy clusters. 
In the standard scenario of inflation, these primordial density fluctuations originate from scalar quantum  fluctuations during the inflationary epoch. 
These fluctuations 
 are first stretched to super-horizon scales during inflation and then re-enter the horizon in the form of classical (but random) density perturbations.
Furthermore, inflation also causes the amplification of tensor perturbations of the spacetime metric, resulting in the generation of primordial gravitational waves (GWs). These waves are referred to as inflationary gravitational waves (IGWs) and represent a key gravitational wave signal originating from the early universe.
\begin{figure}[!ht]
	\centering
\includegraphics[width=0.495\textwidth]{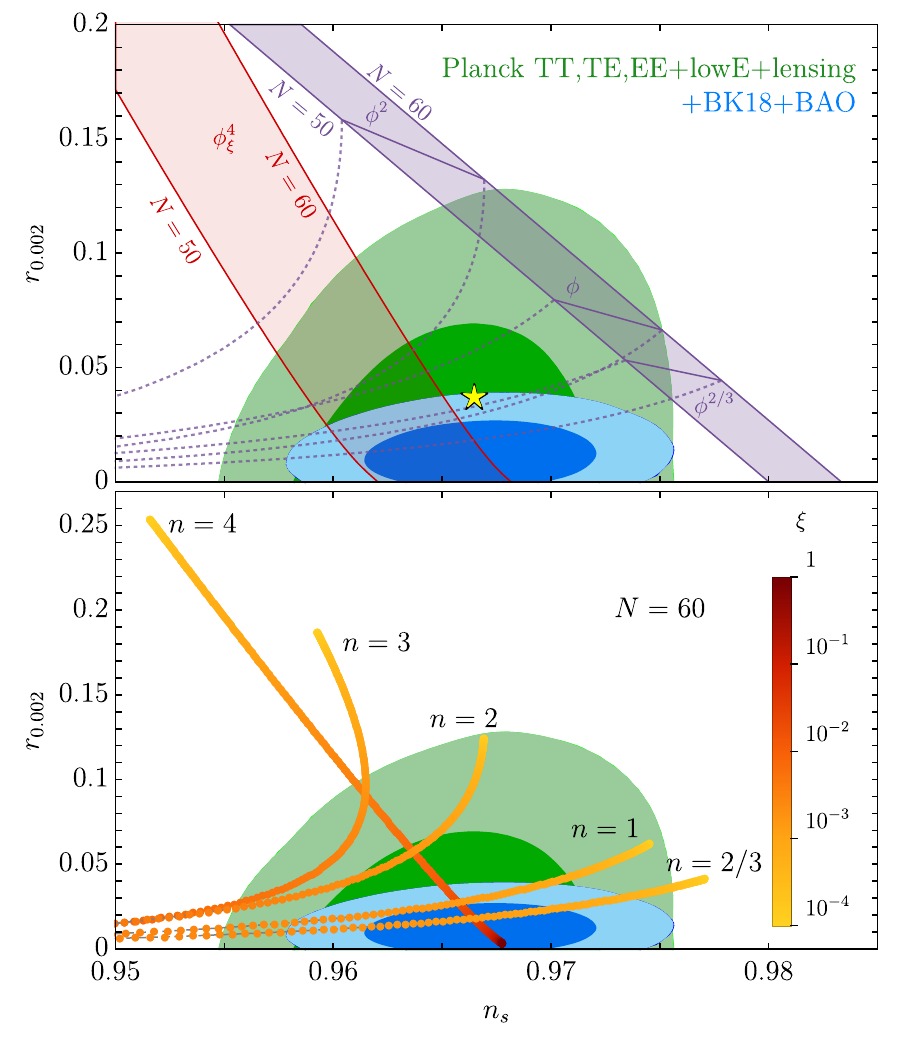}
\vspace{-0.5cm}
	\caption{
Constraints in the $r$ vs. $n_s$ plane using the Planck
2018 baseline analysis (green regions) and 
 including BICEP/Keck and BAO data (blue regions), cf. ref.\,\cite{BICEP:2021xfz}. 
 We show the 65\% and 95\% confidence contours (cf. \href{http://bicepkeck.org/bk18_2021_release.html}{BK18 Data Products}). 
 We confront the experimental constraints with the theoretical prediction of chaotic inflationary models with and without non-minimal coupling to gravity. 
 {\it Top panel.} The  
 lavender band corresponds to the values of $(n_s,r)$ given by eq.\,(\ref{eq:ChaosPred}) with $\xi = 0$ (and $50\leqslant N \leqslant 60$).  
 Inside the lavender band, we highlight the cases with $n=2,1,2/3$. The dotted lines in lavender show the trajectories given by $\xi >0$ for $n=2,1,2/3$ and $N=50,60$.  
 The red band on the left 
 corresponds to the case with $n=4$ and $\xi > 0$. 
 The yellow star is the prediction of the model studied in 
 section\,\ref{sec:Model}.
  {\it Bottom panel.} 
 We consider $N=60$ and $n=4,3,2,1,2/3$. The legend shows the values of $\xi$ as we follow the trajectories given by the non-minimally coupled chaotic models. 
 }
\label{fig:CMBConstraint}
\end{figure}

In this context, the tensor-to-scalar ratio $r$ is a fundamental parameter used to characterize the magnitude of IGWs relative to scalar density fluctuations generated during the inflationary epoch. It quantifies the strength of the GW signal compared to the density perturbations responsible for structure formation; 
similarly, the spectral index $n_s$ is another essential parameter in inflationary cosmology, and it describes the scale dependence of the primordial density fluctuations generated during inflation.  
Technically speaking, $n_s$ and $r$ enter the expressions for the power spectra of scalar and tensor perturbations -- respectively, $P_{\zeta}(k)$ and $P_{t}(k)$ -- via
\begin{align}
P_{\zeta}(k) = A_s\left(
\frac{k}{k_{\star}}
\right)^{n_s-1}\,,~~~
P_{t}(k) = r A_s\left(
\frac{k}{k_{\star}}
\right)^{n_{T}}\,,\label{eq:PowerSpactra}
\end{align}
where $A_s$ is the amplitude of
primordial scalar perturbations at the so-called pivot scale 
$k_{\star}$ and $n_T$ is the tensor spectral index.
We remark that single-field slow-roll models of inflation satisfy the so-called consistency relation $n_T =
-r/8$\,\cite{Copeland:1993ie}, leading to the expectation that
the tensor spectrum should be slightly red-tilted (that is, $n_T < 0$).
For ease of reading, in the above expression for $P_{\zeta}(k)$ we neglect the possible running of the spectral index (however, we will include it in our analysis in section\,\ref{sec:Model}).  
$k$ is the comoving wavenumber of the scalar and tensor perturbations in Fourier space. 
The scales relevant for Cosmic
Microwave Background (CMB) observations are of order $10^{-3} \lesssim 
k\,\,[\textrm{Mpc}^{-1}] \lesssim 2\times 10^{-1}$\,\,\cite{Planck:2018vyg}. 
Throughout this work, we set the pivot scale to the value 
$k_{\star} = 0.05$ Mpc$^{-1}$. 

In the following, we explore the implications that experimental constraints on $r$ and $n_s$ impose on several prototype inflationary models. While it cannot be claimed that this discussion is original, it is beneficial to delve into it as it will provide us with insights into the necessary ingredients that an inflationary model must possess to ensure compatibility with experimental data.

The recent BICEP/Keck constraints on $r$ and $n_s$ present significant evidence against various well-known inflationary models\,\cite{BICEP:2021xfz} (BK18 hereafter). 
These include natural inflation, models featuring monomial potentials, and the Coleman-Weinberg potentials. 
The experimental constraints are summarized in fig.\,\ref{fig:CMBConstraint} (see caption for details).
For definiteness, consider monomial  
chaotic inflation models characterized by the scalar potential 
\begin{align}
V(\phi) = \frac{g^{n-2}}{M^{n-4}}\phi^n\,,\label{eq:Chaos}
\end{align}
where $g$ and $M$ represent some fundamental coupling and mass scale and $\phi$ is the inflaton field that 
drives, under slow-roll conditions, the accelerated background expansion of the primordial universe.  
Eq.\,(\ref{eq:Chaos}) predicts
\begin{align}
n_s = 1 - \frac{2(2+n)}{n+4N}\,,~~~~
r = \frac{16 n}{n+4N}\,,
\label{eq:ChaosPred}
\end{align}
where $N$ is the number of $e$-folds in between the time at which the pivot scale $k_{\star}$ crosses the Hubble horizon and the end of inflation.
Eq.\,(\ref{eq:ChaosPred}) corresponds to the diagonal band shaded in dark lavender in the top panel of fig.\,\ref{fig:CMBConstraint} with $50 \leqslant N \leqslant 60$. 
We highlight the models with 
$n=2,1,2/3$. All these models are now disfavoured by the comparison with BK18 data. Already at this stage, therefore, some modifications over the vanilla models in eq.\,(\ref{eq:ChaosPred}) are required to get back in line with the observational constraints.  

A simple and motivated possibility consists in
the Lagrangian density 
\begin{align}
\frac{\mathcal{L}}{\sqrt{-\mathrm{g}}} = 
-\frac{\MPl^2}{2}\big[1+f(\phi)\big]R 
+\frac{\mathrm{g}_{\mu\nu}}{2}
(\partial^{\mu}\phi)(\partial^{\nu}\phi) -
V(\phi)\,,\label{eq:NMC}
\end{align}
where $\mathrm{g}_{\mu\nu}$ is the space-time metric with determinant $\mathrm{g}$, $R$ the Ricci scalar and $f(\phi)$ represents a non-minimal coupling between gravity and the inflaton field\,\cite{Spokoiny:1984bd,Lucchin:1985ip,Salopek:1988qh,Fakir:1990eg}. 
In eq.\,(\ref{eq:NMC}) the reduced Planck mass is $\MPl = \sqrt{1/8\pi G_N} \approx 2.4\times 10^{18}$ GeV (with $G_N$ the Newton's gravitational constant).
We consider the functional form 
$f(\phi) = \xi\phi^2/\MPl^2$ with $\xi$ dubbed non-minimal coupling. 
In the course of this work, we shall only consider values of 
the non-minimal coupling $\xi \geqslant 0$.
If we consider the limit of small non-minimal coupling and large $e$-fold number, $\xi \ll 1$ and $1/N \ll 1$, 
we find the compact analytical approximations 
\begin{align}
n_s & = 1 - \frac{2(2+n)}{n+4N}
-\frac{\xi}{2}(24 - 14 n + n^2) 
+ O(\xi^2,N^{-2})\,,\\
r & = \frac{16n}{n+4N} 
+4n\xi(-12 + n)+ O(\xi^2,N^{-2})\,.
\end{align}
First, we consider the tensor-to-scalar ratio. We observe that, unless one takes large values $n>12$, the correction to $r$ is always negative. 
The presence of the non-minimal coupling, therefore, tends to lower the value of $r$ with respect to chaotic models. 
We then move to consider the spectral index. Interestingly, 
we observe that the correction to $n_s$ is positive (negative) for $2 < p < 12$ ($p<2 \lor p>12$). 
The presence of the non-minimal coupling, therefore, tends to shift towards larger (smaller) values the prediction of $n_s$ that we get in chaotic models if $p>2$ ($p<2$). 
The above intuition finds confirmation in the top panel of fig.\,\ref{fig:CMBConstraint} in which we compute $r$ and $n_s$ numerically without relying on any approximation on $\xi$ and $N$. 
For definiteness, we consider the case with $n=2,1,2/3$. 
In the $r$ vs. $n_s$ plane, the prediction of chaotic inflation moves towards smaller $r$ and smaller $n_s$, as expected, for increasing  values of $\xi$.
For instance, we observe that the model with 
$n = 2/3$ becomes 
compatible with the K18 data for 
$\xi \simeq 10^{-3}$ (and $N=60$ $e$-fold of inflation). 
In the bottom panel of fig.\,\ref{fig:CMBConstraint}, we plot in the $(n_s,r)$ plane the model trajectories for $N=60$ as we change the non-minimal coupling $\xi \geqslant 0$. We consider the cases 
with $n=4,3,2,1,2/3$. 
Along each curve, the 
color difference is associated to different values of $\xi$ according to the enclosed legend.
As far as the cases with $n>2$ are concerned, we focus in particular on the (well-known) case with $n=4$. 
This case is highlighted in red in 
fig.\,\ref{fig:CMBConstraint}.
The prediction for the tensor-to-scalar ratio of chaotic inflation (not even visible on the $y$-axis scale of the figure) becomes smaller for increasing $\xi$ while, in agreement with our preceding argument, the value of $n_s$ increases accordingly. All in all, the model finds perfect agreement with BK18 data for $\xi \gtrsim 5\times 10^{-2}$ as shown in the bottom panel of fig.\,\ref{fig:CMBConstraint}.

We now turn our attention to the recently detected signal of GWs at the nHz frequency and its implications concerning inflationary models.
 The recent data released 
 by several international Pulsar Timing Array (PTA) collaborations --
 NANOGrav\,\cite{NANOGrav:2023gor,NANOGrav:2023hfp,NANOGrav:2023hvm,NANOGrav:2023hde},  
 EPTA and InPTA\,\cite{Antoniadis:2023rey,Antoniadis:2023utw,Antoniadis:2023zhi}, PPTA\,\cite{Reardon:2023gzh,Zic:2023gta,Reardon:2023zen} and
    CPTA\,\cite{Xu:2023wog} -- provided strong evidence of the Hellings-Downs pattern in the angular correlations of pulsar timing residuals. This pattern is the trademark of the presence of GWs. Among the collaborations,  the 15-year pulsar-timing data set collected by the NANOGrav collaboration (NANOGrav\,15 hereafter) is particularly noteworthy for its stringent constraints and the largest statistical evidence of the Hellings-Downs pattern. 
    
The estimated amplitude and spectrum of the gravitational-wave background align with what we would expect from a population of supermassive black-hole binaries (SMBHB), supporting astrophysical predictions. However, it is important to note that the possibility of more unusual cosmological and astrophysical sources cannot be ruled out entirely. 
In ref.\,\cite{NANOGrav:2023hvm}, the NANOGrav collaboration  
explored potential cosmological explanations for the observed GW signal. Specifically, cosmic inflation, scalar-induced gravitational waves, first-order phase transitions, cosmic strings, and domain walls were considered as possible interpretations. 
Remarkably, all these models, except for stable cosmic strings of field theory origin, can reproduce the observed signal. When compared to the conventional interpretation involving SMBHB, many cosmological models appear to offer a better fit, with Bayes factors ranging from 10 to 100. However, it is crucial to emphasize that these outcomes heavily rely on modeling assumptions concerning the cosmic SMBHB population. Therefore, to date, they should not be considered as conclusive evidence for new physics.

In this paper, we are interested in the inflationary interpretation of the NANOGrav\,15 signal. 
Assuming the latter to be a power-law, the signal corresponds to a fraction energy density in GW today ($\Omega_{\textrm{GW}}$) as described by the following functional form\,\cite{Allen:1997ad}
\begin{align}
\Omega_{\textrm{GW}}(f) \equiv 
\frac{1}{\rho_{\textrm{cr}}}
\frac{d\rho_{\textrm{GW}}}{d\log f}
=  
\frac{2\pi^2 A^2 f_{\textrm{yr}}^2}{
3H_0^2
}\left(
\frac{f}{f_{\textrm{yr}}}
\right)^{\gamma}\,,\label{eq:OmegaPar}
\end{align}
where $d\rho_{\textrm{GW}}$ is the energy density of the gravitational radiation contained in the frequency
range $[f,f+df]$, $\rho_{\textrm{cr}} = 3H_0^2/8\pi G_N$ is the critical energy density, 
$H_0 = h \times 100$ km/s/Mpc the current Hubble rate (we adopt $h=0.674$ in our analysis) and $f_{\textrm{yr}} \equiv (1\,\textrm{yr})^{-1} \simeq 
32$ nHz. $A$ and $\gamma$ are the amplitude and spectral tilt of the power-law, respectively.
The analysis of ref.\,\cite{NANOGrav:2023gor} favors the spectral tilt $\gamma = 2.2^{+0.9}_{-0.8}$  where the errors are quoted at the 2-$\sigma$ level and we consider the fit done in ref.\,\cite{Cannizzaro:2023mgc} including the first 9 frequency bins.

At first glance, 
the simplest option is to relate the NANOGrav\,15 signal directly with 
inflationary tensor perturbations. 
Assuming standard single-field slow-roll inflation, however, one finds that  the IGW amplitude on
PTA scales is 
undetectably small with $A \lesssim 10^{-18}$. 
This is because, as previously discussed,  
the amplitude of the power spectrum of tensor perturbations is bounded from above by current CMB observations, $r \lesssim  0.036$ at 95\% confidence level, and the spectral index $n_T$ in eq.\,(\ref{eq:PowerSpactra}) is  given by the so-called
consistency relation, $n_T = -r/8$\,\cite{Copeland:1993ie}.
This result seems to suggest that explaining the NANOGrav\,15 signal with IGWs requires a  strong violation of 
the consistency relation. 
Such possibility has been discussed in refs.\,\cite{Vagnozzi:2020gtf,NANOGrav:2023hvm,Vagnozzi:2023lwo,Jiang:2023gfe,Ben-Dayan:2023lwd} and, although it has not been definitively excluded, it appears to be highly unlikely. Phenomenologically, it requires the combination of very large $n_T \simeq 1.8 \pm 0.3$ (cf. ref.\,\cite{Choudhury:2023kam} for a quintessential interpretation of such blue tilted tensor spectrum)
and very low reheating temperature $T_{\textrm{rh}} \lesssim 10$ MeV\,\cite{Vagnozzi:2023lwo}.
Theoretically, 
within well-motivated inflationary models it is challenging to attain such a large value of $n_T$; models that do achieve this goal often predict significant non-Gaussianities, which are ruled out by observations\,\cite{Vagnozzi:2023lwo}.

For the remainder of this study, we will explore an alternative, and in our view more promising, approach.\footnote{
Cf., e.g., refs.\,\cite{Ashoorioon:2022raz,Niu:2023bsr,Basilakos:2023xof,Choudhury:2023hvf} for other explanations related to multi-field inflationary models, no-scale Supergravity and Galileon inflation.
}

The underlying idea is to consider the second-order tensor perturbations that are generated by scalar perturbations\,\cite{Tomita:1975kj,Matarrese:1993zf,Acquaviva:2002ud,Mollerach:2003nq,Ananda:2006af,Baumann:2007zm} (rather than the first-order tensor perturbations, i.e., those discussed earlier). 
These second-order tensor perturbations give rise to the so-called {\it scalar-induced GW} signal. The idea itself is not new. It is well-known that in the presence of inflationary dynamics violating the slow-roll conditions, it is possible to enhance scalar perturbations 
and, consequently, increase second-order tensor perturbations and generate a sizable signal of scalar-induced GWs. 
Not surprisingly, scalar-induced GWs have been indeed proposed as a potential explanation for the NANOGrav\,15 signal (cf., e.g., refs.\,\cite{NANOGrav:2023hvm,Madge:2023cak,Balaji:2023ehk}). However, implementing this mechanism within a motivated inflationary model that is in agreement with experimental CMB bounds is an aspect that has been relatively unexplored thus far. 
The reason is not difficult to grasp. In single-field inflation models, incorporating a dynamics that violates the slow-roll conditions and produces a sizable signal of scalar-induced GWs usually worsens the agreement with experimental CMB data (cf. ref.\,\cite{Madge:2023cak} for a recent discussion). The task of finding models that are not excluded becomes highly non-trivial, especially considering, as discussed in this introduction, the challenging constraints imposed by BK18. 
One possibility is to introduce a semi-analytical modeling of the inflationary dynamics through a series of phases characterized by constant 
values of the second Hubble parameter. 
This approach (pursued in ref.\,\cite{Firouzjahi:2023lzg} as far as the comparison with NANOGrav\,15 is concerned) was dubbed reverse engineering approach in ref.\,\cite{Franciolini:2022pav} (see also ref.\,\cite{Byrnes:2018txb}). 
This approach has the advantage of providing a clear description of the inflationary dynamics in terms of a handful of parameters directly related to physical observables. However, the drawback is that it cannot be directly connected to explicit quantum field theory (QFT) models since it provides only a numerical reconstruction of the inflationary potential. 
In the following, we shall focus on a more traditional QFT approach.

We structure the rest of our work as follows. 
In section\,\ref{sec:Model},  building upon the intuition 	strengthen in this introduction,  we introduce and discuss our theoretical model.  
In section\,\ref{sec:Data}, we compute the relevant observables and compare our results with experimental data. 
In particular, in section\,\ref{sec:PS} we compute the classical dynamics and the power spectra of scalar and tensor perturbations; in section\,\ref{sec:Tech} and section\,\ref{sec:ABU} we compute the abundance of primordial black holes (PBHs); 
in section\,\ref{sec:SIWG} we compute the scalar-induced GW signal; in section\,\ref{sec:comments} we further comment about our theoretical setting.
In section\,\ref{sec:Discu}, 
we critically examine the comparison between the model's prediction in terms of the scalar-induced GWs signal and the recent experimental measurement of NANOGrav\,15.
Finally, we sum up and conclude.

\section{The model}\label{sec:Model}

We consider the effective QFT described, in curved space, by the following Lagrangian density 
\begin{align}
\frac{\mathcal{L}}{\sqrt{-\mathrm{g}}} & = 
\bigg[
-\frac{\MPl}{2}
\left(1 + 
\frac{\xi\phi^2}{
\MPl^2
}
\right)R 
+\frac{1}{2}
(\partial^{\mu}\phi)
(\partial_{\mu}\phi)
- V(\phi)
\bigg]\,,\label{eq:Lagra}\\
V(\phi) & = 
\sum_{n=2}^{\mathcal{N}}
a_n\frac{g^{n-2}}{M^{n-4}}\phi^n = \frac{M^4}{g^2}
\sum_{n=2}^{\mathcal{N}} a_n\left(\frac{g\phi}{M}\right)^n\,,\label{eq:Pote}
\end{align}
where $a_n$ are genuine dimensionless numbers while $g$ and $M$ represent some coupling and mass. Notice that the combination $g\phi/M$ in eq.\,(\ref{eq:Pote}) is dimensionless too. 

Before proceeding, several observations are warranted. 
First of all, eq.\,(\ref{eq:Lagra}) includes the non-minimal coupling with gravity. As discussed in section\,\ref{sec:Intro}, the presence of this term can play a crucial role in comparing the model predictions with the stringent observational constraints imposed by the 
CMB.\footnote{However, employing a non-minimal coupling is not the only possible choice. For instance, another option is presented by the so-called $\alpha$-attractor models, as shown in ref.\,\cite{Kallosh:2022ggf,Braglia:2022phb}. 
Instead, we refer to ref.\,\cite{HosseiniMansoori:2023mqh} for a study within the context of $\mathbb{T}^2$ inflation.} 
From the QFT perspective, the presence of a non-minimal coupling is entirely expected. In fact, this term does not violate any of the symmetries of the theory, and it is therefore anticipated to appear in the Lagrangian density (and if it were absent, it would still be generated radiatively at the level of the effective action). 
Secondly, we observe how, in the spirit of an effective theory, we are allowing both renormalizable terms ($n\leqslant 4$) and non-renormalizable effective operators ($n>4$) to appear in the scalar potential.\footnote{ 
Note also that we do not consider effective operators involving derivatives of the scalar field. The presence of these terms could represent an interesting option to investigate in possible follow-ups to our analysis.} The basic assumption is that the sector that has been integrated out to obtain the effective Lagrangian in eq.\,(\ref{eq:Lagra}) can be entirely characterized by a single coupling ($g$) and a single mass scale ($M$).
As far as these two parameters are concerned, an educated assumption would be to posit the relationship
\begin{align}
g\MPl = M\,,\label{eq:String}
\end{align}
that correctly identifies $\MPl$ as a scale defined by the ratio between a mass and a coupling. 
To fix ideas, we can think about $g$ and $M$ as the 
string mass and coupling in the context of a stringy ultraviolet completion of gravity.  
We then rewrite the potential in the form
\begin{align}
V(\phi) = M^2\MPl^2
\bigg\{ &
a_2\left(\frac{\phi}{\MPl}\right)^2 + 
a_3\left(\frac{\phi}{\MPl}\right)^3 + \nn\\
&
a_4\left(\frac{\phi}{\MPl}\right)^4\bigg[
1 + \sum_{n=5}^{\mathcal{N}}
\frac{a_n}{a_4}
\left(\frac{\phi}{\MPl}\right)^{n-4}
\bigg]\bigg\}\,,\label{eq:PotV}
\end{align}
in which we factor out from the curly bracket the dependence on dimensionful quantities. It is now clear that we can work with field values in units of the reduced Planck mass.

After a Weyl transformation, the theory in the Einstein frame takes the following form
\begin{align}
\frac{\mathcal{L}}{\sqrt{-\mathrm{g}}}  = 
-\frac{\MPl^2}{2}R + 
\frac{1}{2}(\partial_{\mu}\varphi)(\partial^{\mu}\varphi) - 
\underbrace{\frac{V(\phi)}{(1+\xi\phi^2/\MPl^2)^2}}_{\equiv \tilde{V}(\phi)}\,,\label{eq:EinsteinFrame}
\end{align}
where the canonically normalized scalar field $\varphi$ is related to $\phi$ via the differential relation
\begin{align}
\frac{d\varphi}{d\phi} = 
\frac{1}{(1+\xi\phi^2/\MPl^2)}
\sqrt{1+(1+6\xi)\xi\phi^2/\MPl^2}\,.
\end{align}
At this point, the second important lesson that we have learned from the discussion in the introduction comes into play: We need the inflationary dynamics to produce an enhancement of scalar perturbations (and, consequently, also of tensor perturbations as a second-order effect). In this study, we choose to employ the presence of a quasi-stationary inflection point in the potential\,\cite{Ivanov:1994pa,Garcia-Bellido:2017mdw,Germani:2017bcs,Ballesteros:2017fsr,Ballesteros:2020qam}. 
In this case, the period of time during which we have a violation of the slow-roll conditions is dubbed ultra slow-roll (USR) phase.
Having a quasi-stationary inflection point in the potential introduces some degree of parametric tuning (cf. ref.\,\cite{Cole:2023wyx} for a comprehensive discussion). However, this is a common feature shared by all models exhibiting a USR phase capable of producing PBHs.

In order to establish a quasi-stationary inflection point in the potential, one possible approach -- as pursued, for instance, in the analysis of refs.\,\cite{Ballesteros:2017fsr,Ballesteros:2020qam} -- involves imposing the conditions $\left.d\tilde{V}/d\phi\right|_{\phi_0} = 0$ and $\left.d^2\tilde{V}/d\phi^2\right|_{\phi_0} = 0$ (that define an exact stationary inflection point at field value $\phi_0$) directly on the potential $\tilde{V}(\phi)$, yielding a system of equations solvable in terms of $a_{2,3}$.
Since what is actually required is an approximately stationary inflection point rather than an exact one\,\cite{Ballesteros:2017fsr}, two parameters $c_{2,3}$ are subsequently introduced to parameterize small deviations from the previously imposed conditions $\left.d\tilde{V}/d\phi\right|_{\phi_0} = 0$ and $\left.d^2\tilde{V}/d\phi^2\right|_{\phi_0} = 0$.

This approach instills a certain sense of artificiality into the ultimate form of the potential and runs the risk of obfuscating the power counting of the effective field theory. In this work, hence, we opt to pursue a more direct path: rather simply, we retain the structure of the potential as depicted in eq.\,(\ref{eq:PotV}) and seek an approximately stationary inflection point by explicitly tuning the coefficients $a_{2,3}$.\footnote{Of course, nothing prevents one from employing the approach of ref.\,\cite{Ballesteros:2020qam} and subsequently deriving the values of the coefficients $a_{2,3}$ by working backward from the values of $c_{2,3}$ and the equations that were used to remove the explicit dependence on $a_{2,3}$ in the potential.}

\begin{figure}[!t]
	\centering
\includegraphics[width=0.495\textwidth]{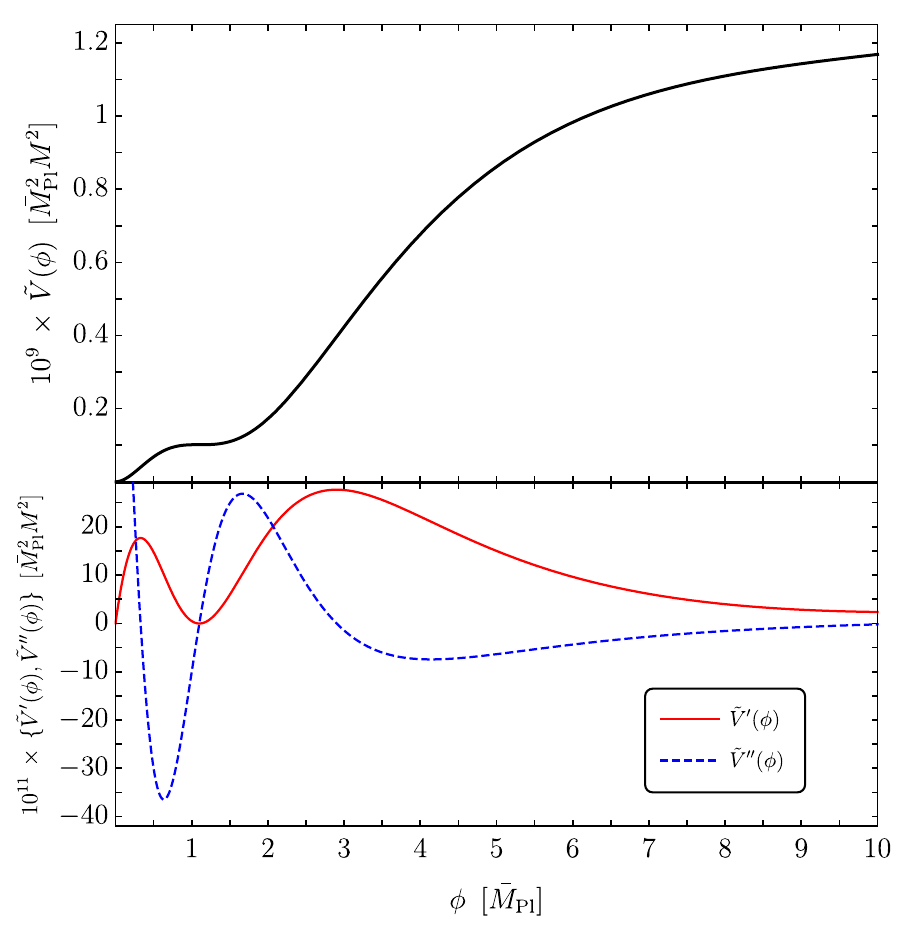}
	\caption{
 Top panel.
 Einstein-frame potential $\tilde{V}(\phi)$, cf. eqs.\,(\ref{eq:PotV},\ref{eq:EinsteinFrame}), for the realization of our model given in table\,\ref{tab:Parameters}.
 Bottom panel. We show the derivatives $\tilde{V}^{\prime} = d\tilde{V}/d\phi$ e 
 $\tilde{V}^{\prime\prime} = d^2\tilde{V}/d\phi^2$.  An approximate stationary inflection point $\tilde{V}^{\prime} = \tilde{V}^{\prime\prime} \simeq 0$ emerges at field values around $\phi \simeq 1$.  
 }
\label{fig:InfPot}
\end{figure}

With regards to the value of $\mathcal{N}$, we find that the minimum value required to adequately fulfill the objectives of our analysis is $\mathcal{N} = 6$, and it is this value that we shall focus on henceforth. 
In conclusion, therefore, the potential we are considering takes the following form
\begin{align}
\tilde{V}(\phi) = \frac{M^2\MPl^2 a_4}{
(1 + \xi\phi^2)^2
}\left(
\tilde{a}_2 \phi^2 + \tilde{a}_3 \phi^3 + \phi^4  + \tilde{a}_5 \phi^5 + 
\tilde{a}_6 \phi^6
\right)\,,
\end{align}
where $\tilde{a}_i \equiv a_i/a_4$ and field values are understood in units of $\MPl$.
The typical shape of the resulting potential $\tilde{V}(\phi)$ is shown in fig.\,\ref{fig:InfPot}. 
We are left with the following free parameter:  $\tilde{a}_{2,3}$ (that control the quasi-stationary inflection point), the non-minimal coupling $\xi$, the coefficients $\tilde{a}_{5,6}$ (that control the relative size of the non-renormalizable operators), and
 the overall dimensionless coefficient 
$a_4$ (that controls the normalization of the potential $\tilde{V}(\phi)$ in units of the dimensionful combination $M^2 \MPl^2$). 

In the end, we arrived at a form of the inflationary potential which is a combination of the pieces discussed individually in the introduction, but now tightly integrated into an effective QFT.

\section{Comparison with data}\label{sec:Data}

In this section, we confront theory with data. 
In section\,\ref{sec:PS}, we compute the power spectra of scalar and tensor perturbations given the model discussed in section\,\ref{sec:Model}. 
In section\,\ref{sec:Tech} and section\,\ref{sec:ABU}, we compute the abundance of PBHs.
In section\,\ref{sec:SIWG}, we compute the signal of scalar-induced GWs. Finally, in section\,\ref{sec:comments} we make further comments about our theoretical setup.

We will strive to set up our discussion in a general manner, but the results of various analyses and corresponding figures will be directly presented for the best realization of our model. For this latter case, the values of the relevant parameters are shown in table\,\ref{tab:Parameters}.
\begin{table}[htp]
	\begin{center}
		\begin{adjustbox}{max width=.485\textwidth}
		\begin{tabular}{||c||c|c|c|c|c|c||}\hline
     Parameter &  $\xi$ & $\tilde{a}_2$ & $\tilde{a}_3$ & $\tilde{a}_5$ & $\tilde{a}_6$ & $a_4 M^2/\MPl^2$ \\ \hline
      Value & 
      $0.28$ & 
      $2.18$ &
      $-2.49$ & 
      $-6.22\times 10^{-2}$  & 
      $2.20\times 10^{-3}$   & 
      $2.66\times 10^{-10}$
      \\ \hline\hline
      \end{tabular}
	\end{adjustbox}
	\end{center}\vspace{-0.25cm}\caption{
 Numerical values (with two digits precision) for the benchmark realization of our model that is explicitly studied in the course of this work. 
    }\label{tab:Parameters}
\end{table}

\subsection{Classical dynamics and power spectra}\label{sec:PS}

\begin{figure}[!t]
	\centering
\includegraphics[width=0.495\textwidth]{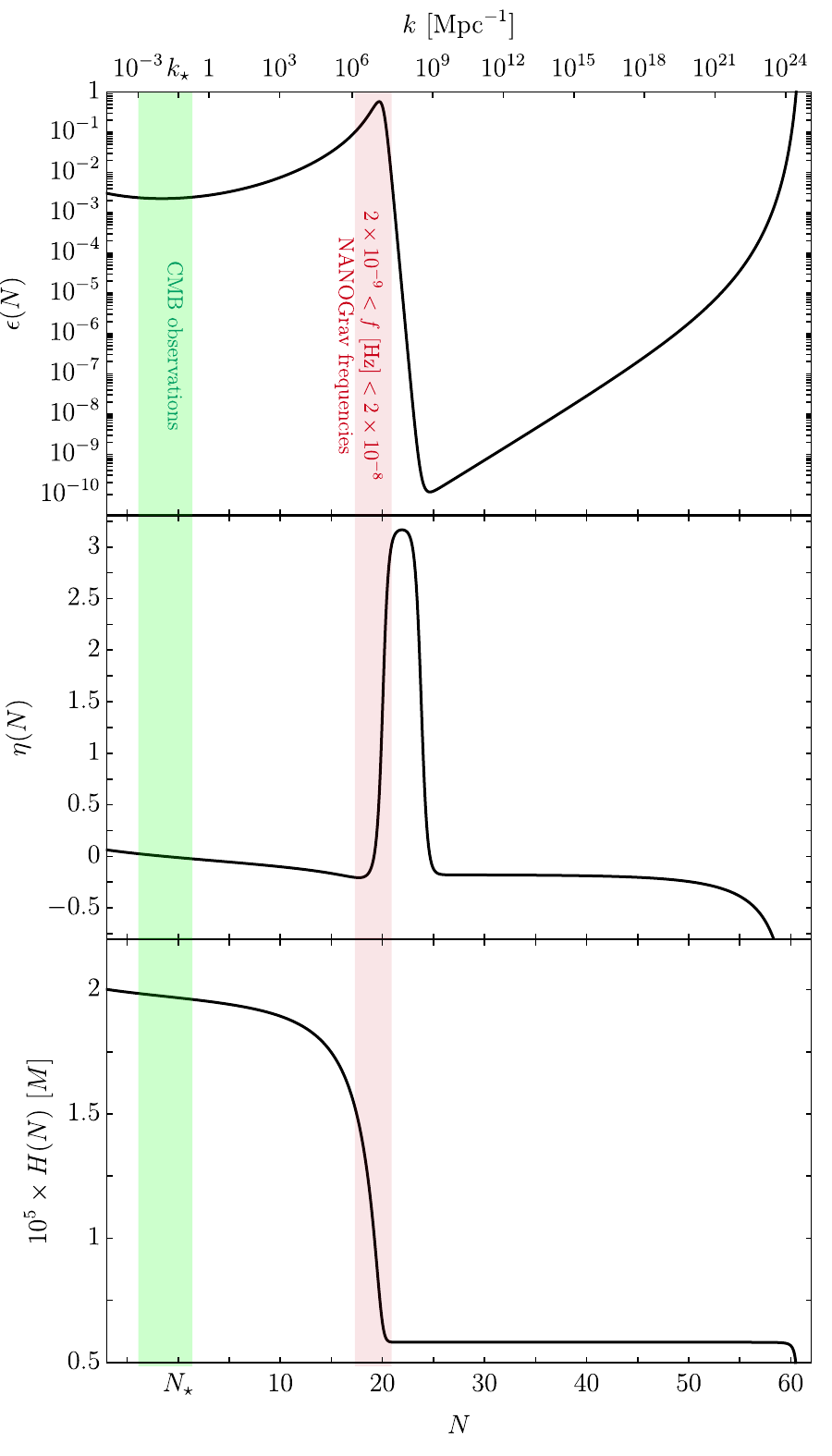}
	\caption{	
 Classical dynamics. From top to bottom we show, as a function of the $e$-fold number, the Hubble parameters $\epsilon$ and $\eta$ and the Hubble rate $H$.
 }
\label{fig:ClassicalDyn}
\end{figure}

First of all, we solve the background dynamics of the inflaton field. The latter is described by the equation of motion
\begin{align}\label{eq:EoM}
\varphi^{\prime\prime} + 3\varphi^{\prime} - \frac{1}{2}\left(\varphi^{\prime}\right)^3 
+\left[
3-\frac{1}{2}\left(\varphi^{\prime}\right)^2
\right]\frac{d\log U}{d\varphi}=0\,,
\end{align}
where  $\varphi = \varphi(N)$ and $^{\prime}$ indicates derivative with respect to the  number of  $e$-folds.
$U(\varphi)\equiv \tilde{V}[\phi(\varphi)]$ is the potential felt by the canonically normalized inflaton field.  
In fig.\,\ref{fig:ClassicalDyn}, we display the time evolution of some background quantities of particular relevance for our discussion (please note that these figures, as well as all others presented in this section, already correspond to the best realization of the model discussed in this analysis). 
We focus on the time evolution of the Hubble rate $H=\dot{a}/a$ (bottom panel; $a$ is the scale factor of the Friedmann-Lema\^{\i}tre-Robertson-Walker space-time with
$\dot{} \equiv d/dt$ that indicates the  derivative with respect to cosmic time $t$ with $dN/dt=H$) and the Hubble parameters
$\epsilon = -\dot{H}/H^2$ and 
$\eta \equiv -\ddot{H}/2H\dot{H}$ (top and central panel, respectively). 
We require that $60$ $e$-folds of accelerated expansion occur between the moment when the pivot scale $k_{\star} = 0.05$ Mpc$^{-1}$ exits the comoving Hubble horizon -- at time $N_{\star}$ defined by $k_{\star} = a(N_{\star})H(N_{\star})$ -- and the end of inflation; we conventionally normalize the $e$-fold time in such a way that $N_{\star} = 0$. 
We solve eq.\,(\ref{eq:EoM}) using the following initial conditions at $e$-fold time $N_{\star} = 0$. 
As far as $\varphi(N_{\star}) \equiv \varphi_{\star}$ is 
concerned, we use the field value such that the spectral index in slow-roll approximation -- that is,
$n_s = 1 + 2\eta_{U} - 6\epsilon_U$ -- takes a value in agreement with CMB observations. 
In the previous expression $\eta_{U}$ and  $\epsilon_U$ are the potential slow-roll parameters 
\begin{align}
\epsilon_U = \frac{\MPl^2}{2}\left(\frac{1}{U}\frac{dU}{d\varphi}\right)^2\,,~~~~~
\eta_U = \frac{\MPl^2}{U}
\frac{d^2 U}{d\varphi^2}\,.
\end{align}
For definiteness, at this stage of the analysis we require $n_s = 0.960$. 
Once $\varphi_\star$ is found, the initial condition for the field velocity is given by the slow-roll condition $\varphi^{\prime}(N_{\star}) = -\sqrt{2\epsilon_U(\varphi_{\star})}$.

We would like to emphasize some key points that need to be kept in mind for the rest of the analysis.

The presence of a quasi-stationary inflection point, as expected, introduces a phase of USR in the inflationary dynamics (formally defined by the condition $\eta > 3/2$, which in our model's realization that is explicitly shown in this section occurs in the range $21 \lesssim N \lesssim 23$). During this phase, the inflaton's velocity decreases to nearly negligible values (though not exactly zero). Consequently, this explains the behavior of $\epsilon$ (as the latter essentially measures the inflaton's velocity). 

The perturbation modes that cross the horizon immediately before or during the USR phase are those that are enhanced the most\,\cite{Ballesteros:2020qam,Franciolini:2022pav}. To guide the eye, in fig.\,\ref{fig:ClassicalDyn} we show on the top $x$-axis the values of $k$ that, for each corresponding $N$ on the bottom $x$-axis, solve the equation $k=a(N)H(N)$. 
The rule of thumb  is that the USR  phase must occur in such a way that the modes crossing the comoving Hubble horizon precisely during this phase have comoving wavenumbers $k$ corresponding to typical PTA frequencies (that is, in the nHz range). 
Given the relation between comoving wavenumber $k$ and 
frequency $f$, that is
\begin{align}
f  \simeq 1.55\left(
\frac{k}{10^6\,\textrm{Mpc}^{-1}}
\right)\,\textrm{nHz}\,,
\end{align}
and the frequency range of the NANOGrav\,15 signal,  
$2\times 10^{-9} \lesssim 
f\,\,[\textrm{Hz}]
\lesssim 2\times 10^{-8}$,
it turns out that the USR must take place at around $\log(10^7/0.05) \approx 20$ $e$-folds after the pivot scale $k_{\star}$ has left the horizon. 
The interval of $k$ corresponding to the  
NANOGrav\,15 signal is shaded in light crimson in fig.\,\ref{fig:ClassicalDyn}.

It is important to emphasize that in order to match the PTA frequencies, the USR phase must occur much earlier than is commonly encountered in single-field inflation models in which the presence of a quasi-stationary inflection point is utilized to produce asteroid-mass primordial black holes (PBHs) which could potentially account for the entirety of dark matter (in such case, one typically needs $10^{13} \lesssim k\,[\textrm{Mpc}^{-1}] \lesssim 10^{14}$; cf. the analysis in ref.\,\cite{Ballesteros:2017fsr,Ballesteros:2020qam} for the case of  the polynomial model).

In the slow-roll approximation, the scalar and tensor power spectra in 
eq.\,(\ref{eq:PowerSpactra}) can be computed as
\begin{align}
P_{\zeta}(k) = \frac{H(N_k)^2}{8\pi^2\MPl^2\epsilon(N_k)}\,,~~~~~~
P_t(k) = \frac{2H(N_k)^2}{\pi^2\MPl^2}\,. \label{eq:PSSR}
\end{align}
Here, it is implied that the dependence on $N$ appearing in the right-hand side is transformed into the dependence on $k$ using the horizon crossing condition $k=a(N_k)H(N_k)$. 
Since $H$ is almost constant, 
eq.\,(\ref{eq:PSSR}) implies that if just reverse the time dependence of $\epsilon$ we can get an idea about the behavior of the power spectrum of scalar perturbations.  
During the USR phase, the value of $\epsilon$ decreases exponentially fast, and consequently, we expect to observe a rapid growth in the power spectrum of scalar perturbations. 

The region shaded in green in fig.\,\ref{fig:ClassicalDyn} corresponds to the interval of $k$ constrained by CMB measurement. 
The amplitude $A_s$ of the scalar power spectrum at the pivot scale il very well constrained, and we use the value 
$A_s = 2.1\times 10^{-9}$\,\cite{Planck:2018vyg}.  Eq.\,(\ref{eq:PSSR}) in combination with the Einstein equation $(3-\epsilon)H^2 = U/\MPl^2$ gives 
\begin{align}
A_s = \frac{U(\varphi_{\star})}{
8\pi^2\MPl^4 \epsilon(N_{\star})
[3-\epsilon(N_{\star})]
}\,.
\end{align}
By means of this equation, we choose the coefficient $a_4M^2/\MPl^2$ of the scalar potential in such a way as to be compatible with  
$A_s = 2.1\times 10^{-9}$.

As far as the comparison with the other CMB constraints is concerned, we compute the scalar power spectrum beyond  the slow-roll approximation used in eq.\,(\ref{eq:PSSR}).  
To this end, we numerically solve the 
Mukhanov-Sasaki (MS) equation\,\cite{Mukhanov:1988jd,Sasaki:1986hm}
\begin{align}\label{eq:M-S}
&
u_k^{\prime\prime} + 
 (1-\epsilon)u_k^{\prime}+ \nn\\
&
\left[
\frac{k^2}{(aH)^2} + (1+\epsilon-\eta)(\eta - 2) - (\epsilon - \eta)^{\prime}
\right]u_k = 0\,,
\end{align} 
with sub-horizon Bunch-Davies initial conditions\,\cite{Bunch:1978yq} at $N \ll N_k$, where $N_k$ indicates the horizon crossing time for the mode $k$, that is the time at which we have $k = a(N_k)H(N_k)$. 
Eq.\,(\ref{eq:M-S}) is written with the number of $e$-fold as time parameter.
The scalar power spectrum $P_{\zeta}(k)$ is then given by 
\begin{align}\label{eq:PS}
P_{\zeta}(k) = \frac{k^3}{2\pi^2}\left|\frac{u_k(N)}{a(N)\varphi^{\prime}(N)}\right|^2_{N > N_{\rm F}(k)}\,
\end{align}
The power spectrum $P_{\zeta}(k)$ does not depend on time because the meaning of 
eq.\,(\ref{eq:PS}) is that $P_{\zeta}(k)$ must be evaluated 
after the time $N_{\rm F}(k)$ at which the mode $|u_k(N)/z(N)|$ freezes to the constant value that is conserved until its 
horizon re-entry. 
We then have $N_{\rm F}(k) \equiv {\rm max}\{N_k, N_{\tiny{\textrm{end}}}\}$, where $N_{\tiny{\textrm{end}}}$ corresponds to the  end of the USR phase. 
Modes that exit the horizon {\it before} the time $N_{\tiny{\textrm{end}}}$ (that is modes such that $N_k < N_{\tiny{\textrm{end}}}$) are 
not conserved  (even though super-horizon) because they experience afterward the negative friction phase. Consequently, 
for these modes their contribution to  eq.\,(\ref{eq:PS}) must be evaluated at any time $N > N_{\tiny{\textrm{end}}} > N_k$ after the USR phase ends. Contrariwise, modes 
that exit the horizon {\it after} the time $N_{\tiny{\textrm{end}}}$ (that is modes such that $N_k > N_{\tiny{\textrm{end}}}$) freeze 
to their constant value after they become super-horizon. Consequently, 
as customary, the contribution of these modes to  eq.\,(\ref{eq:PS}) must be evaluated at any time $N > N_k > N_{\tiny{\textrm{end}}}$. 
\begin{figure}[!t]
	\centering
\includegraphics[width=0.495\textwidth]{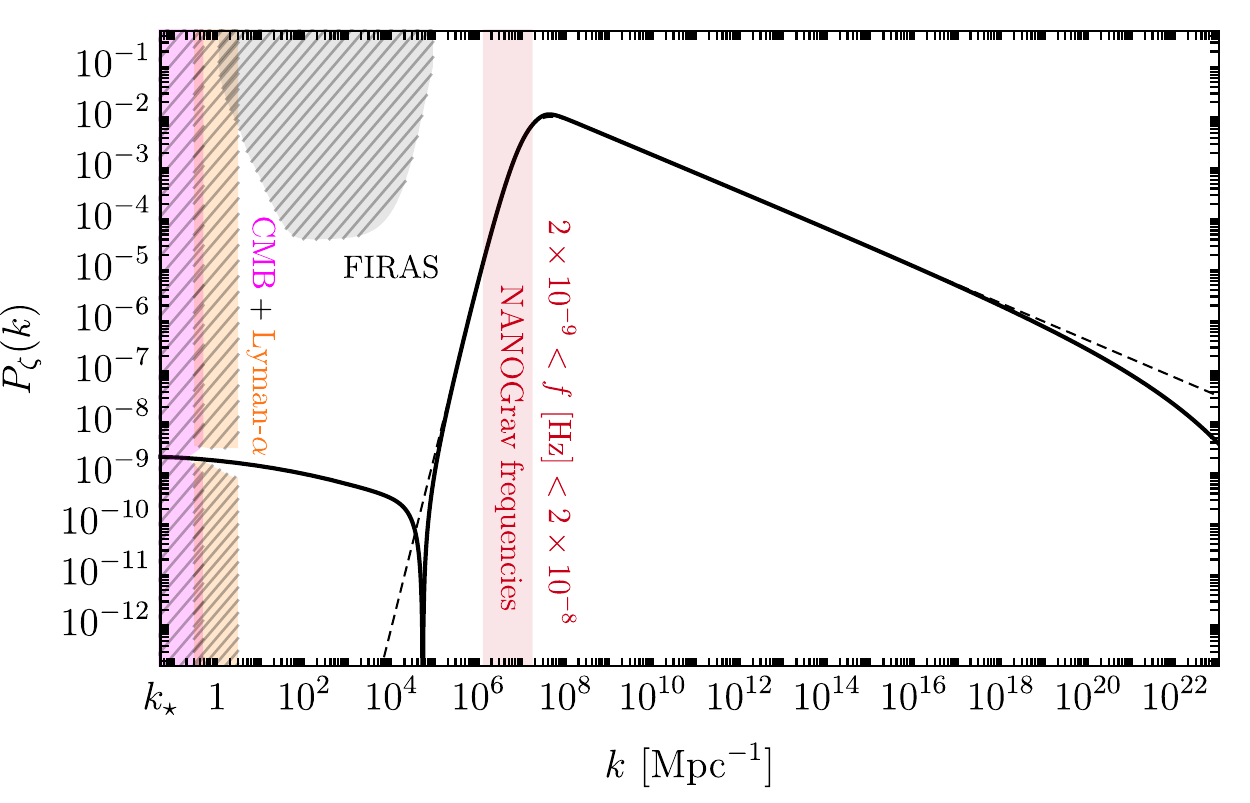}
	\caption{
Power spectrum of scalar perturbations $P_{\zeta}(k)$ as a function of the comoving wavenumber $k$ computed numerically by solving the MS equation. We plot the region excluded by CMB anisotropy measurements, ref.\,\cite{Planck:2018jri}, the FIRAS bound on CMB spectral distortions, ref.\,\cite{Chluba:2012we} (see also ref.\,\cite{Jeong:2014gna}) and the bound obtained from Lyman-$\alpha$ forest data \cite{Bird:2010mp}.  
 }
\label{fig:PowerSpectrum}
\end{figure}
Our  result is shown in fig.\,\ref{fig:PowerSpectrum} where we also plot the  most relevant constraints 
from CMB anisotropy measurements, CMB spectral distortions, and  Lyman-$\alpha$ forest data (see caption for details). 

The most relevant aspect that we would like to comment on is the following. 
As inferred from the analysis based on the slow-roll approximation, the scalar power spectrum is characterized by a peak located around the scales $k$ relevant for PTA measurements. 
The requirement for the peak to be at this position leads to a kind of compression of the power spectrum at large scales (i.e. those relevant for CMB observations).
\begin{figure}[!t]
	\centering
\includegraphics[width=0.495\textwidth]{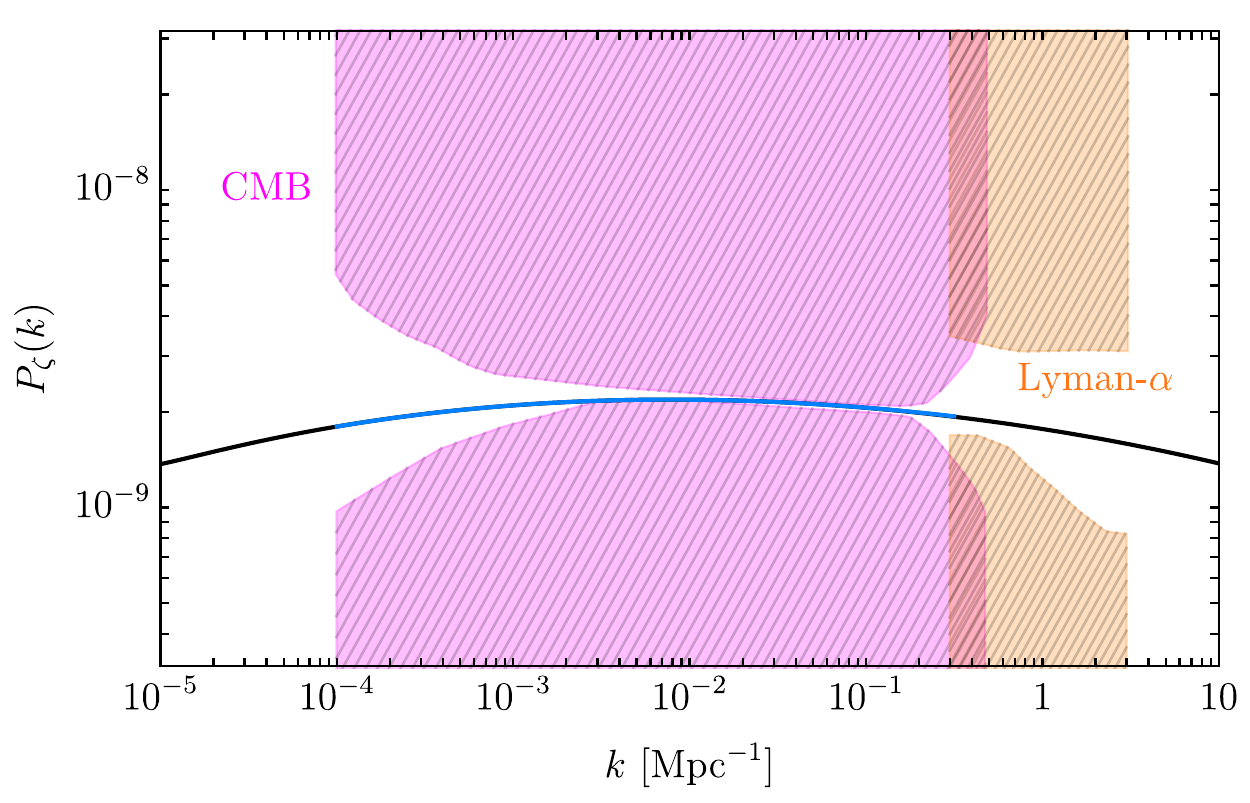}
	\caption{
 Same as in fig.\,\ref{fig:PowerSpectrum} but zoomed in at  large scales relevant for CMB and Lyman-$\alpha$ observations. 
 The black solid line 
 corresponds to the numerical solution of the M-S equation. The superimposed  azure line 
 corresponds to the best-fit function obtained in the observational window 
 $10^{-4} \leqslant k\,\,[\textrm{Mpc}^{-1}]\leqslant  4\times 10^{-1}$.
 }
\label{fig:PowerZoom}
\end{figure}
This point is well illustrated in fig.\,\ref{fig:PowerZoom}, where we show a zoom-in on the scales relevant for CMB observations. It is evident how the model produces a power spectrum with a significant amount of running. 
 In order to make contact with CMB observables,  at scales $10^{-4} \leqslant k\,\,[\textrm{Mpc}^{-1}]\leqslant  4\times 10^{-1}$ we fit our power spectrum against the parametric function\,\cite{Planck:2018vyg} (same as in eq.\,(\ref{eq:PowerSpactra}) but now including the effect of the running of the spectral index)
\begin{align}\label{eq:ParametricPS}
P_{\zeta}(k) =
A_s\left(\frac{k}{k_{\star}}\right)^{
{n_s - 1 +\frac{\alpha}{2}\log\frac{k}{k_*}+
\frac{\vartheta}{6}\log^2\frac{k}{k_*}+\dots}
}\,,
\end{align}      
with $\alpha = dn_s/d\log k$, $\vartheta= d^2n_s/d\log k^2$.
At the pivot scale $k_* = 0.05$ Mpc$^{-1}$, we find
\begin{align}
A_s \simeq 2.12\times 10^{-9}\,,~ 
n_s = 0.966\,,~\alpha =  -0.017\,,~\vartheta = 10^{-3}\,.
\end{align}
These values are in good agreement with Planck's data\,\cite{Planck:2018vyg}.  
As previously anticipated, we observe a non-negligible running in the spectral index, which nonetheless remains compatible with experimental bounds. For comparison, we quote (at the 65\% confidence level) the value of the running parameter $\alpha$ (treated as a single additional parameter to the base $\Lambda$CDM model), $\alpha = -0.0045 \pm 0.0067$\,\cite{Planck:2018vyg}. Our value of $\alpha$ in eq.\,(\ref{eq:ParametricPS}) is negative and within the 65\%  confidence interval. For completeness, we find $n_{s,0.002} \simeq 0.996$  (consistent at the 65\% confidence level with the experimental value $n_{s,0.002} = 0.979 \pm 0.021$\,\cite{Planck:2018vyg}).

The value $n_s = 0.966$ represents the $x$-coordinate of the yellow star in fig.\,\ref{fig:CMBConstraint}.  
The function in eq.\,(\ref{eq:ParametricPS}) for our best-fit values of its parameters is shown as an azure curve in fig.\,\ref{fig:PowerZoom}.

Finally, we compute the tensor-to-scalar ratio at the reference scale $k=0.002$ Mpc$^{-1}$ which is relevant for the comparison with BK18. 
We find
\begin{align}
r_{0.002}  = 16\,\epsilon(N_{0.002}) \simeq 0.037\,,
\end{align}
where $N_{0.002}$ is the $e$-fold time at which  the reference scale $k=0.002$ Mpc$^{-1}$ crosses the comoving Hubble horizon.  
The value $r_{0.002} = 0.037$ represents the $y$-coordinate of the yellow star in fig.\,\ref{fig:CMBConstraint}. 

As an additional piece of information, we notice that around the peak the power spectrum in fig.\,\ref{fig:PowerSpectrum} is well approximated by a broken power law described by the functional form
\begin{align}
P_{\zeta}^{\textrm{\tiny{BPL}}}(k)  = 
\frac{
A(\alpha+\beta)^{\gamma}
}{
\big[\beta\big(\frac{k}{x k_{\textrm{peak}}}\big)^{-\alpha/\gamma} 
+ \alpha(\frac{k}{x k_{\textrm{peak}}}\big)^{\beta/\gamma}\big]^{\gamma}
}\,,\label{eq:BPL}
\end{align}
with best-fit parameters 
\begin{align}
A\simeq 10^{-2}\,,~~
\alpha \simeq 3.42\,,~~
\beta \simeq 0.37\,,~~\gamma \simeq 2.46\,,~~
x\simeq 1.12\,.
\end{align}
Eq.\,(\ref{eq:BPL}) corresponds to the dashed line in fig.\,\ref{fig:PowerSpectrum}.

\subsection{From the curvature power spectrum to the abundance of primordial black holes}\label{sec:Tech}

This section serves as a technical intermediary, bridging the computation of the power spectrum on super-horizon scales during inflation 
to the estimate of the PBH abundance (elaborated in section\,\ref{sec:ABU}). 

In a nutshell, the physical picture is as follows. In the previous section, we computed the power spectrum of curvature perturbations on super-horizon scales during inflation. Once these perturbations become constant, quantum fluctuations undergo a process of classicalization\,\cite{Kiefer:1998jk}, and the curvature field transitions into a classical random field whose statistics, in the Gaussian approximation, is entirely determined by the power 
spectrum (which quantifies the contribution to the variance of the curvature perturbation field per logarithmic $k$-bin). 

After the end of inflation, the universe enters the radiation era. The comoving Hubble radius initiates its expansion, causing the initially super-horizon modes to inevitably evolve towards their horizon re-entry. 
Curvature perturbations generated during inflation are imprinted onto the primordial plasma, creating 
regions with a slightly higher and slightly lower density of matter. 
These overdensities and underdensities start to grow under the influence of gravitational attraction.  
The over-dense regions -- provided they exceed a certain threshold -- can eventually undergo gravitational collapse, resulting in the formation of PBHs. This is the essence of the so-called {\it standard scenario of PBH formation}, in which PBHs are formed through the collapse of overdensities of inflationary origin during the evolution of the post-inflationary(and typically radiation-dominated) early universe\,\cite{Zeldovich:1967lct,Hawking:1971ei,Carr:1974nx}. 

In order to ensure self-consistency throughout our work, let us establish clear definitions for the notation we will be using.
The relevant metric, in the
uniform density slicing, is defined through the line element $ds^2 = dt^2 - a(t)^2 e^{2\zeta(r)}(dr^2 + r^2d\Omega^2)$; 
$\zeta(r)$ is the conserved comoving curvature perturbation random field defined on
super-horizon scales, and written assuming spherical symmetry. 
The line element $ds^2$ describes  a locally perturbed region that would eventually collapse to form a PBH; 
such regions are very rare, and spherical symmetry represents a good approximation.\footnote{Cf. refs.\,\cite{Kuhnel:2016exn,Akrami:2016vrq} for a discussion about 
non-spherical collapse.} 
The areal radius is defined by
$R(r,t) \equiv a(t)re^{\zeta(r)}$ where $r$ is the comoving radial coordinate 
that appears in the expression of the metric given above.  
The compaction function is defined as twice the mass excess over the areal radius (as for the rest of this section, we set $G_N = 1$)
\begin{align}
\mathcal{C}(r,t) \equiv 
\frac{
2\big[
M(r,t) - M_b(r,t)
\big]
}{R(r,t)}\,,   
\end{align}
where $M(r,t)$ is the Misner-Sharp mass within a sphere of radius $R(r,t)$
and $M_b(r,t)$ is the background mass within the same areal radius calculated
with respect to the background energy density. We have 
$M_b(r,t) = \rho_b(t)V_b(r,t)$ with $V_b(r,t) = (4\pi/3)R(r,t)^3$; 
$\rho_b(t)$ is the average background radiation energy density, $\rho_b(t) = 3H(t)^2/8\pi$, while $\delta\rho(\vec{x},t) \equiv \rho(\vec{x},t)- \rho_b(t)$ its perturbation. The so-called density contrast field is defined  as $\delta(\vec{x},t) \equiv \delta\rho(\vec{x},t)/\rho_b(t)$.
Equivalently, therefore, the compaction function can be written as 
\begin{align}\label{eq:DefinitionCompaction}
\mathcal{C}(r,t) &  = 
\frac{2}{R(r,t)}\int_{S^2_R} d^{3}\vec{x}
\left[\rho(\vec{x},t) - \rho_b(t)\right] \nn\\
& = 
\frac{2}{R(r,t)}
\underbrace{\int_{S^2_R} d^{3}\vec{x}\,\rho_b(t)\,\delta(\vec{x},t)}_{\coloneqq\,\delta M(r,t)}\,.
\end{align}
On super-horizon scales and in the so-called  gradient expansion 
approximation\,\cite{Shibata:1999zs},
 the density contrast is related to the comoving perturbation field by 
 means of\,\cite{Harada:2015yda}
\begin{align}\label{eq:SphericalDelta}
&\delta(r,t) =\nn\\
&-\frac{2(1+\omega)}{5+3\omega}\left(\frac{1}{aH}\right)^2 
e^{-2\zeta(r)}\left[
\zeta^{\prime\prime}(r) + \frac{2}{r}\zeta^{\prime}(r) + \frac{1}{2}\zeta^{\prime}(r)^2
\right]\,,
\end{align}
where we assume spherical symmetry. 
$\omega \equiv p/\rho$ is the equation of state for a perfect fluid and $p$ is the pressure;
$\omega = 1/3$ during the radiation-dominated era (but away from the QCD phase transition during which the equation of state becomes softer). 
If we plug in the above expression into 
eq.\,(\ref{eq:DefinitionCompaction}), the compaction function takes the form
\begin{align}\label{eq:CompactionFull}
\mathcal{C}(r,t) = 
-\frac{6(1+\omega)}{5+3\omega}\,r\,\zeta^{\prime}(r)\left[
1 + \frac{r}{2}\zeta^{\prime}(r)
\right]\equiv \mathcal{C}(r)\,,
\end{align}
where the last redefinition means that under the assumptions underlying eq.\,(\ref{eq:SphericalDelta}) the compaction function becomes time-independent.

The comoving length scale $r_m$ is defined as the scale at which the compaction function is maximized. 
The condition 
$\mathcal{C}^{\prime}(r_m) = 0$,
 by means of eq.\,(\ref{eq:CompactionFull}), reads
\begin{align}
\zeta^{\prime}(r_m) + r_m\zeta^{\prime\prime}(r_m) = 0\,.  \label{eq:ZetaCri}  
\end{align} 
Consider now the mass excess averaged within a spherical region of areal radius $R$. 
This quantity, that we indicate with the ratio
$\delta M(r,t)/M_b(r,t)$, is given by the following definition
\begin{align}
\frac{\delta M(r,t)}{M_b(r,t)} \equiv \frac{1}{V_b(r,t)}
\int_{S^2_R} d^{3}\vec{x}\,\delta\rho(\vec{x},t)\,.\label{eq:AvMassEx}
\end{align}
Furthermore, we define the horizon crossing time $t_m$ as the time at which the condition
\begin{align}
a(t_m)H(t_m) r_m e^{\zeta(r_m)} = 1\,,\label{eq:HorizonCross}
\end{align}
is met. Given the definition of  the areal radius, the physical scale corresponding to the maximum of the compaction function is given by 
$R_m \equiv R(r_m,t_m)  = a(t_m)r_m e^{\zeta(r_m)}$. Eq.\,(\ref{eq:HorizonCross}) tells that at time $t_m$ the physical scale $R_m$ is crossing the cosmological horizon $1/H(t_m)$.  
If the value of the compaction function at its maximum is larger then a certain threshold, then very soon after the crossing time $t_m$  
the parcel of space enclosed in the region defined by the comoving radius $r_m e^{\zeta(r_m)}$ undergoes a gravitational collapse.

Notice that the scale $\tilde{r}_m \equiv r_m e^{\zeta(r_m)}$ is  the
typical scale of a collapsing perturbation in real space. In Fourier space, modes with comoving wavenumber $k = O(\tilde{r}_m^{-1})$ are those that will contribute the most to the process of PBH formation.  

The relevance of these definition is that it is possible to derive the fundamental relation\,\cite{Musco:2018rwt} 
\begin{align}
\delta_m \coloneqq \frac{\delta M(r_m,t_m)}{M_b(r_m,t_m)} \overset{(a)}{=} \mathcal{C}(r_m) 
\overset{(b)}{=} 3\delta(r_m,t_m)\,.
\label{eq:MainCompa}
\end{align} 
The first equality {\it (a)} follows 
using eq.\,(\ref{eq:SphericalDelta}), together with the condition in eq.(\ref{eq:HorizonCross}), inside eq.\,(\ref{eq:AvMassEx}). 
The second equality {\it (b)} follows again from eq.\,(\ref{eq:SphericalDelta})  evaluated at horizon crossing
together with the condition $\zeta^{\prime}(r_m) + r_m\zeta^{\prime\prime}(r_m) = 0$ that defines $r_m$. 
Eq.\,(\ref{eq:MainCompa}) implies that the compaction function at its maximum equals the density contrast volume-averaged, at  horizon crossing time $t_m$, over a spherical region whose size is set by $r_m$. Consequently, 
the criterion\,\cite{Shibata:1999zs} according to which a PBH forms if $\mathcal{C}(r_m)$
is larger than some 
critical threshold $\delta_{\textrm{th}}$ can be equally well formulated in terms of the volume-averaged density contrast. 
The latter, therefore, becomes a crucial quantity in the theoretical estimate of the PBH abundance.

In the subsequent analysis, our primary focus lies in calculating the threshold parameter $\delta_{\textrm{th}}$, which we shall adhere to as prescribed in 
ref.\,\cite{Musco:2020jjb}.

First, we define 
$P_{\zeta}^{\mathcal{T}}(k,\tau) \equiv P_{\zeta}(k)\mathcal{T}(k\tau)^2$ where 
$\mathcal{T}(k\tau)$ is the linear radiation transfer function  defined by\,\cite{Blais:2002gw}
\begin{align}
\mathcal{T}(y) \equiv   
3\left[
\frac{
\sin(c_s y) - 
(c_s y)\cos(c_s y)
}{
(c_s  y)^3
}
\right]\,,\label{eq:Transfer}
\end{align}
with $\tau  = 1/aH$ and $c_s = 1/\sqrt{3}$ the sound speed of the relativistic fluid. 
$\mathcal{T}(k\tau)$ introduces a time-dependence on the power spectrum (which is otherwise  time-independent since it was computed at super-horizon scales) and has the effect of smoothing out the modes as they re-enter the horizon during the radiation epoch (notice that $\lim_{kc_s\tau\gg 1}\mathcal{T}(k\tau) = 0$ meaning that sub-horizon modes with $k\gg  (c_s\tau)^{-1}$ are smoothed away). 
Physically, the transfer function plays the role of pressure gradients during the collapse.  
In fact, pressure effects act as a smoothing and naturally damp perturbations on scales smaller than the sound horizon  $c_s\tau = c_s/aH$.
\begin{figure}[!t]
	\centering
\includegraphics[width=0.495\textwidth]{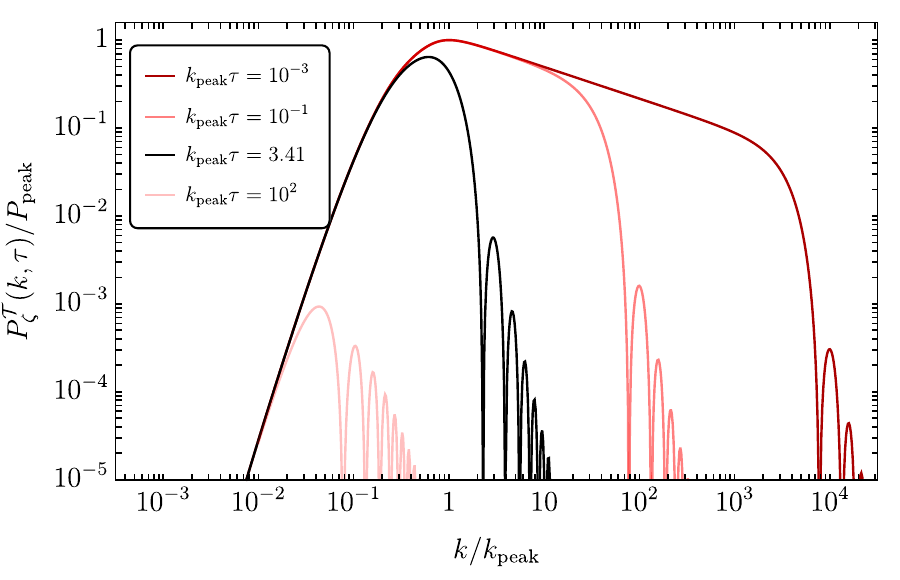}
	\caption{
We plot $P_{\zeta}^{\mathcal{T}}(k,\tau)$ (normalized at the peak value $P_{\textrm{peak}}\equiv P_{\zeta}(k_{\textrm{peak}})$ of the time-independent power spectrum) at different increasing $\tau$ (from right to left, cf. the plot legend). 
 }
\label{fig:SmoothedPS}
\end{figure}

For illustration, we plot $P_{\zeta}^{\mathcal{T}}(k,\tau)$ at different $\tau$ (in units of $1/k_{\textrm{peak}}$) in fig.\,\ref{fig:SmoothedPS} focusing on the peak region of the power spectrum located at comoving wavenumber $k_{\textrm{peak}}$. 
As time passes, modes of increasingly lower frequency become sub-horizon, and their corresponding contribution to the power spectrum gets suppressed by the transfer function.

The comoving length scale $r_m$ solves the equation\,\cite{Musco:2020jjb}
\begin{align}
\int
\frac{dk}{k} \left[
(k^2 r_m^2  - 1)
\frac{\sin(k r_m)}{k r_m} + 
\cos(k r_m)
\right]
P_{\zeta}^{\mathcal{T}}(k,\tau)
= 0\,. \label{eq:MasterEqRm}
\end{align}
It has been shown in ref.\,\cite{Musco:2020jjb}
 that this integral equation is nothing but a reformulation of eq.\,(\ref{eq:ZetaCri}). 
Formally, the solution to eq.\,(\ref{eq:MasterEqRm}) depends on $\tau$.
\begin{figure}[!t]
	\centering
\includegraphics[width=0.495\textwidth]{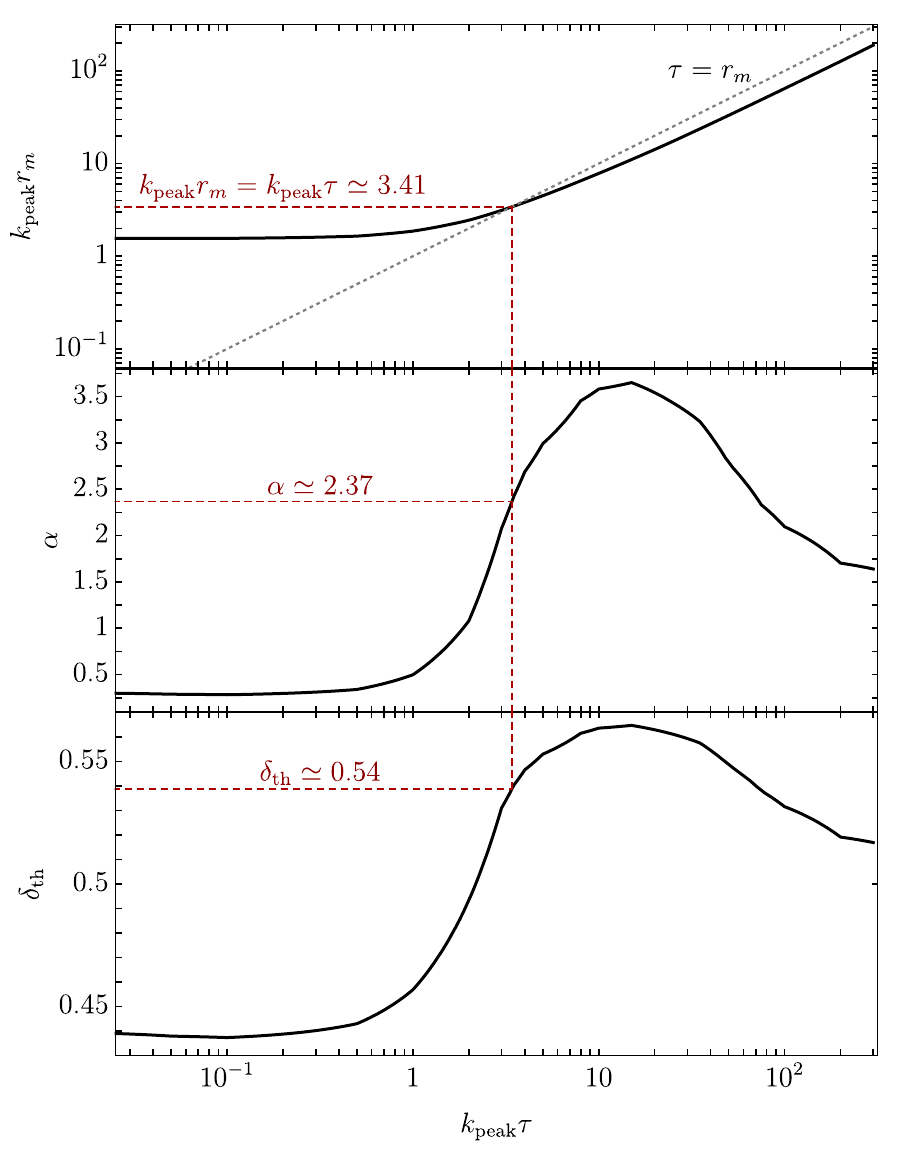}
	\caption{
 Top panel.
Numerical solution of eq.\,(\ref{eq:MasterEqRm}) that gives the scale $r_m$ for a given time $\tau$.  
Central panel. Shape parameter $\alpha$ obtained solving eq.\,(\ref{eq:AlphaEq}). 
Bottom panel.  Threshold parameter $\delta_{\textrm{th}}$ given by 
eq.\,(\ref{eq:Threshold}). 
In all three panels, we highlight the special case with $\tau = r_m$ that corresponds to $k_{\textrm{peak}}r_m \simeq 3.41$, 
$\alpha \simeq 2.37$ and $\delta_{\textrm{th}} \simeq 0.54$.
 }
\label{fig:SpectrumScales}
\end{figure}
We solve it numerically in full generality and show our solution in the top panel of fig.\,\ref{fig:SpectrumScales}. 

The shape parameter is defined as the value of $\alpha$ that solves the equation\,\cite{Musco:2020jjb}
\begin{align}
F(\alpha)[1+&F(\alpha)]\alpha =\nn\\
&
-\frac{1}{2}\left[
1+\frac{
r_m\int dk k \cos(kr_m)P_{\zeta}^{\mathcal{T}}(k,\tau)
}{
\int dk \sin(k r_m)P_{\zeta}^{\mathcal{T}}(k,\tau)
}
\right]\,,\label{eq:AlphaEq}
\end{align}
with  
\begin{align}
F(\alpha) =  \sqrt{
1-\frac{2}{5}e^{-1/\alpha}\frac{
\alpha^{1-5/2\alpha}
}{
\Gamma(5/2\alpha) - 
\Gamma(5/2\alpha,1/\alpha)
}
}\,.
\end{align}
At each time $\tau$, we use the scale $r_m$ obtained by solving eq.\,(\ref{eq:MasterEqRm}).
We plot $\alpha$ in the central panel of fig.\,\ref{fig:SpectrumScales}.
Once the shape parameter $\alpha$ is known we can also compute the threshold parameter $\delta_{\textrm{th}}$ which is given by the analytical formula\,\cite{Escriva:2019phb}
\begin{align}
\delta_{\textrm{th}} =  
\frac{4}{15}
e^{-1/\alpha}\frac{
\alpha^{1-5/2\alpha}
}{
\Gamma(5/2\alpha) - 
\Gamma(5/2\alpha,1/\alpha)
}\,.\label{eq:Threshold}
\end{align}
The parameter $\delta_{\textrm{th}}$ represents the threshold for gravitational collapse (cf. section\,\ref{sec:ABU}). 
The behavior of $\delta_{\textrm{th}}$ is shown in the bottom panel of fig.\,\ref{fig:SpectrumScales}.
It should be noted that the above value is strictly valid in a radiation-dominated universe.

Of particular relevance is the case in which we consider $\tau = r_m$ in 
eq.\,(\ref{eq:MasterEqRm}). 
The reason is twofold. First of all, 
assuming the threshold condition is met, 
the gravitational collapse happens quite quickly soon after the relevant comoving scale $r_m e^{\zeta(r_m)}$ re-enters the horizon\,\cite{Kalaja:2019uju}.
Consequently, we are mostly interested in this particular situation.
Second,  
the horizon crossing condition in eq.\,(\ref{eq:AvMassEx}) implies, at the linear level, $aH r_m = 1$.
With this choice, we find 
\begin{align}
k_{\textrm{peak}}r_m \simeq 3.41\,,  \label{eq:rmNum}
\end{align}
where $k_{\textrm{peak}} \simeq  4\times 10^7$ Mpc$^{-1}$ is the position of the peak of the scalar power spectrum.
As far as the shape parameter is concerned, we find $\alpha = 2.37$.
This value is close to the benchmark value $\alpha = 3$ which characterizes scale invariant and broad power spectra, cf. ref.\,\cite{Musco:2020jjb}.
In  this latter case, one also finds $k_{\textrm{max}}r_m \simeq 4.49$ (to be compared with our eq.\,(\ref{eq:rmNum})) where $k_{\textrm{max}}$ is defined as 
$P_{\zeta}(k) = P_0\Theta(k - k_{\textrm{min}})\Theta(k_{\textrm{max}} - k)$ with $k_{\textrm{max}} \gg k_{\textrm{min}}$ and $\Theta$ the Heaviside step function. 
On the contrary, for a monochromatic curvature spectrum peaked at a given momentum $k_{\delta}$ one finds $k_{\delta}r_m \simeq 2.7$.
As far as the threshold parameter is concerned, we find 
\begin{align}
\delta_{\textrm{th}}\simeq 0.54\,.\label{eq:ThresholdParameter}
\end{align}
The above value of $\delta_{\textrm{th}}$ is again close, at the percent level, to the value $\delta_{\textrm{th}} \simeq 0.55$ that characterizes scale invariant and broad power spectra, cf. ref.\,\cite{Franciolini:2022tfm}. 

The take-home message of this section is that the relevant variable to be used to describe the process of PBH formation is the density contrast $\delta$ volume-averaged over a spherical region of comoving radius $r_m e^{\zeta(r_m)}$ and evaluated at horizon-crossing time $t_m$ defined by $r_m e^{\zeta(r_m)} = 1/a(t_m)H(t_m)$. 
It is in terms of the statistics of this variable that the abundance of PBH will be computed in the next section.
Gravitational collapse takes place if the volume-averaged density contrast at horizon crossing turns out to be larger than the critical value $\delta_{\textrm{th}}$.

Before proceeding, it should be stressed that the value of the threshold computed according to eq.\,(\ref{eq:Threshold}) gives the  threshold at horizon crossing by making a linear extrapolation from the super-horizon regime. 
Ref.\,\cite{Musco:2020jjb} claims that 
the value of the threshold  increases by about a factor
of two if non-linear effects of the higher orders in the
gradient expansion approach are included  (using $\alpha = 2.37$ and their fitting function, we indeed get $\delta_{\textrm{th}}^{\textrm{\tiny{NL}}} \simeq 1.06$). This observation has the potential to be highly significant, and will be further discussed in section\,\ref{sec:Discu}.

Equipped with these insights and findings, we are now poised to calculate the PBH abundance.

\subsection{Abundance of primordial black holes}\label{sec:ABU}

The standard scenario for PBH formation from inflationary overdensities provides a compelling explanation for the origin of PBHs, and it allows for the production of PBHs with masses ranging from microscopic scales to much larger ones. Observational constraints on PBHs, therefore, can provide valuable insights into the early universe and the dynamics of inflation.
We follow the computation detailed in ref.\,\cite{Franciolini:2022pav} (see also ref.\,\cite{Young:2019yug} for previous analysis). 

After the matter-radiation equality, the dark matter fraction consisting of PBHs can be expressed as follows
\begin{align}
\Omega_{\textrm{PBH}} & = 
\int d \log M_H 
\left(
\frac{M_{\rm eq}}{M_H}
\right)^{1/2}\beta(M_H)
\,,\label{eq:Beta}\\
f_{\rm PBH}(M_{\rm PBH}) & = 
\frac{1}{\Omega_{\rm CDM}}\frac{d\Omega_{\rm PBH}}{d\log M_{\rm PBH}}\,,
\end{align}
where $M_H$ is the horizon mass at the time of horizon re-entry, and $M_{\rm eq} \simeq 3 \times 10^{17}\, M_{\odot}$ is the horizon mass at matter-radiation equality. 
We remind that the horizon mass is defined by 
$M_H \equiv (4\pi/3)\rho H^{-3}$ with $\rho$ being the total energy density of the 
universe.  
$\Omega_{\rm CDM}$ represents the cold dark matter density of the Universe, approximately $0.12\,h^{-2}$, with $h = 0.674$ for the Hubble parameter. 

In the standard scenario, the relation between the horizon mass at  time $t$, $M_H(t)$, and the comoving wavenumber
of the sourced primordial curvature perturbation that crosses (re-enters) the horizon at time $t$, $k_H \equiv a(t)H(t)$, 
is given by\,\cite{Kawasaki:2016pql,Sasaki:2018dmp}
\begin{align}
M_H(t) \simeq 17 \times \left(\frac{g_*}{10.75}\right)^{-1/6} \left(\frac{k_H}{10^6\, \textrm{Mpc}^{-1}}\right)^{-2}\, M_{\odot}\,,\label{eq:MHkHRel}
\end{align}
where $g_*$ represents the number of degrees of freedom of relativistic particles and, deep in the radiation epoch, $g_* = 106.75$. The temperature dependence of $g_*$ can be included following ref.\,\cite{Saikawa:2018rcs}, and  converted into a functional dependence on the horizon mass using the relation\,\cite{Byrnes:2018clq} 
\begin{align}
  M_H(T)\simeq 1.5\times 10^5 \left[\frac{g_*(T)}{10.75}\right]^{-1/2}\left(\frac{T}{\textrm{MeV}}\right)^{-2}\, M_{\odot}\,.\label{eq:MHTemp} 
\end{align}.
Eq.\,(\ref{eq:MHkHRel}) simply follows from the definition of $M_H$ and comoving entropy conservation.

The mass of the resulting PBH is given by\,\cite{Choptuik:1992jv,Evans:1994pj,Koike:1995jm,Young:2019yug}:
\begin{align}
M_{\rm PBH} = \mathcal{K} M_H \left[\left(\delta_{\rm L} - \frac{1}{4\Phi}\delta_{\rm L}^2\right) - \delta_{\textrm{th}}\right]^{\gamma}\,.\label{eq:MainNL}
\end{align}
Eq.\,(\ref{eq:MainNL}) accounts for the non-linear relation between curvature perturbations and the density contrast field \cite{DeLuca:2019qsy,Young:2019yug}. More specifically, $\delta_{\rm L}$ represents the linear Gaussian component of the density contrast field, while $\delta_{\textrm{th}}$ is the threshold value for gravitational collapse, referring to the full density contrast field. We computed $\delta_{\textrm{th}}$, assuming a radiation-dominated universe, in eq.\,(\ref{eq:ThresholdParameter}). 
As far as the parameters $\mathcal{K}$, $\gamma$ and $\Phi$ are concerned, we use the numerical values 
$\mathcal{K} = 4.36$, $\gamma = 0.38$ and $\Phi = 2/3$. 
It should be noted that, in full generality, $\mathcal{K}(M_H)$, $\gamma(M_H)$, $\Phi(M_H)$ and $\delta_{\textrm{th}}(M_H)$ are functions of the horizon mass with the values quoted above that 
correspond to the radiation-dominated universe. 
Such dependence on $M_H$ is particularly relevant at the QCD phase transition, during which a softening of the equation of state is observed, leading to a subsequent lowering of the threshold for PBH formation\,\cite{Byrnes:2018clq,Franciolini:2022tfm,Musco:2023dak,Escriva:2022bwe}. This effect is particularly significant when considering models capable of primarily producing solar mass PBHs. 
This is because the scaling in eq.\,(\ref{eq:MHTemp}) implies that the horizon mass at around a temperature 
of a few hundred MeV is $O(1)\,M_{\odot}$. 
On the contrary, our model typically produces sub-solar PBHs. 
Since our scalar power spectrum peaks at $k_{\textrm{peak}} = 4\times 10^7$ Mpc$^{-1}$, eq.\,(\ref{eq:MHkHRel}) implies (if we na\"{\i}vely identify $k_H = k_{\textrm{peak}}$) that we primarily form PBH with sub-solar mass. 

The function $\beta(M_H)$ in eq.\,(\ref{eq:Beta}) is given by
\begin{align}
\beta(M_H) & = 
\int_{\delta_{\textrm{th}}}^{\infty}\frac{M_{\rm PBH}}{M_H}P(\delta)d\delta\,, \label{eq:BetaDef}\\
& = 
\mathcal{K}
\int_{\delta_{\rm L}^{\rm min}}^{\delta_{\rm L}^{\rm max}}
\left(\delta_{\rm L} - \frac{1}{4\Phi}\delta_{\rm L}^2 - \delta_{\textrm{th}}\right)^{\gamma}
P_{\rm G}(\delta_{\rm L})d\delta_{\rm L}\,,\label{eq:GaussianTerm1}\\
P_{\rm G}(\delta_{\rm L}) & = \frac{1}{\sqrt{2\pi}\sigma(M_H)}e^{-\delta_{\rm L}^2/2\sigma^2(M_H)}\,.
\label{eq:GaussianTerm}
\end{align}
The two integration endpoints in eq.\,(\ref{eq:GaussianTerm1}) are 
\begin{align}
\delta_{\rm L}^{\rm min} = 2\Phi\left(
1 - \sqrt{1-\frac{\delta_{\textrm{th}}}{\Phi}}\,
\right)\,,~~~~~~\delta_{\rm L}^{\rm max} = 2\Phi\,.
\end{align}
The variance in eq.\,(\ref{eq:GaussianTerm}) 
can be computed by integrating the power spectrum of scalar perturbations
\begin{align}
\sigma^2(M_H) = \frac{4}{9}\Phi^2
\int_{0}^{\infty}(k R_H)^4W(k \tilde{r}_m)^2 P_{\zeta}^{\mathcal{T}}(k,R_H)
d \log k\,,\label{eq:Sigma}
\end{align}
where $P_{\zeta}^{\mathcal{T}}(k,R_H)  = 
P_{\zeta}(k)\mathcal{T}(k R_H)^2$ has been already defined above eq.\,(\ref{eq:Transfer}).
$R_H = 1/aH$ is the comoving Hubble radius.
$W(y)$ is the Fourier transform of the top-hat window
function 
\begin{align}\label{eq:WeT}
W(y)  = 3\bigg[
\frac{\sin(y) - y\cos(y)}{y^3}\bigg] \,.
\end{align}
Let us further comment about eq.\,(\ref{eq:Sigma}) since it is the quantity that primarily controls the abundance of PBH. 
As we have discussed in section\,\ref{sec:Tech}, the key parameter is the volume-averaged density contrast at horizon crossing during radiation (cf. eq.\,(\ref{eq:MainCompa}) and related discussion).
In eq.\,(\ref{eq:GaussianTerm}), $\sigma$ is the variance of the (Gaussian) linear component of the density contrast. 
In linear approximation, the relation between density contrast and the curvature perturbation field reads $\delta_{\textrm{L}}(\vec{x},t) = 
-(2/3)\Phi(1/aH)^2\triangle\zeta(\vec{x})$. Consequently, the variance of $\delta_{\textrm{L}}$ is 
$\sigma^2 = (4/9)\Phi^2\int_0^{\infty}(k/aH)^4P_{\zeta}(k)d\log k$.  This explains the main structure of eq.\,(\ref{eq:Sigma}). In addition, we have the presence of two filters.

On the one hand, the window function $W (kR)$ 
is responsible for smoothing the density contrast field over a finite volume of size set by the comoving
length scale $R$ (notice that $\lim_{kR\gg 1}W(kR) = 0$ meaning that high-frequency modes with $k\gg  1/R$ are smoothed away).  
Physically, the window function is needed since -- as reviewed in section\,\ref{sec:Tech} -- the key parameter to use for PBH formation  
 is the volume-averaged density contrast. 
 For this reason, the smoothing scale of the window function in eq.\,(\ref{eq:Sigma}) is set equal to $\tilde{r}_m = r_m e^{\zeta(r_m)}$ (that is, $\tilde{r}_m = r_m$ at linear level). 
 
On the other one,
$\mathcal{T}(k\tau)$ -- as already discussed  -- has the effect of smoothing out sub-horizon modes (cf. fig.\,\ref{fig:SmoothedPS} and related discussion) and its characteristic physical scale is the size of the sound horizon, where pressure gradients effectively smooth the perturbations. 
In eq.\,(\ref{eq:Sigma}),  we have $R_H = 1/aH$.

Let us now comment about  the $M_H$-dependence on $\sigma$. 
First, we set equal to two scales $r_m = R_H$ in eq.\,(\ref{eq:Sigma}).  
This choice enforces the horizon-crossing condition in eq.\,(\ref{eq:HorizonCross}) since the volume-averaged density contrast has to be evaluated at horizon crossing.  
For clarity's sake, we rewrite
\begin{align}
\sigma^2(M_H) = \frac{4}{9}\Phi^2
\int_{0}^{\infty}(k r_m)^4W(k r_m)^2 P_{\zeta}^{\mathcal{T}}(k,r_m)
d \log k\,.\label{eq:Sigma2}
\end{align}
We then use eq.\,(\ref{eq:MHkHRel}) to relate the smoothing
scale $r_m$ in eq.\,(\ref{eq:Sigma2}) 
with the horizon mass $M_H$ by means of the identification $k_H = r_m^{-1} = a(t_m)H(t_m)$. 
\begin{figure}[!t]
	\centering
\includegraphics[width=0.495\textwidth]{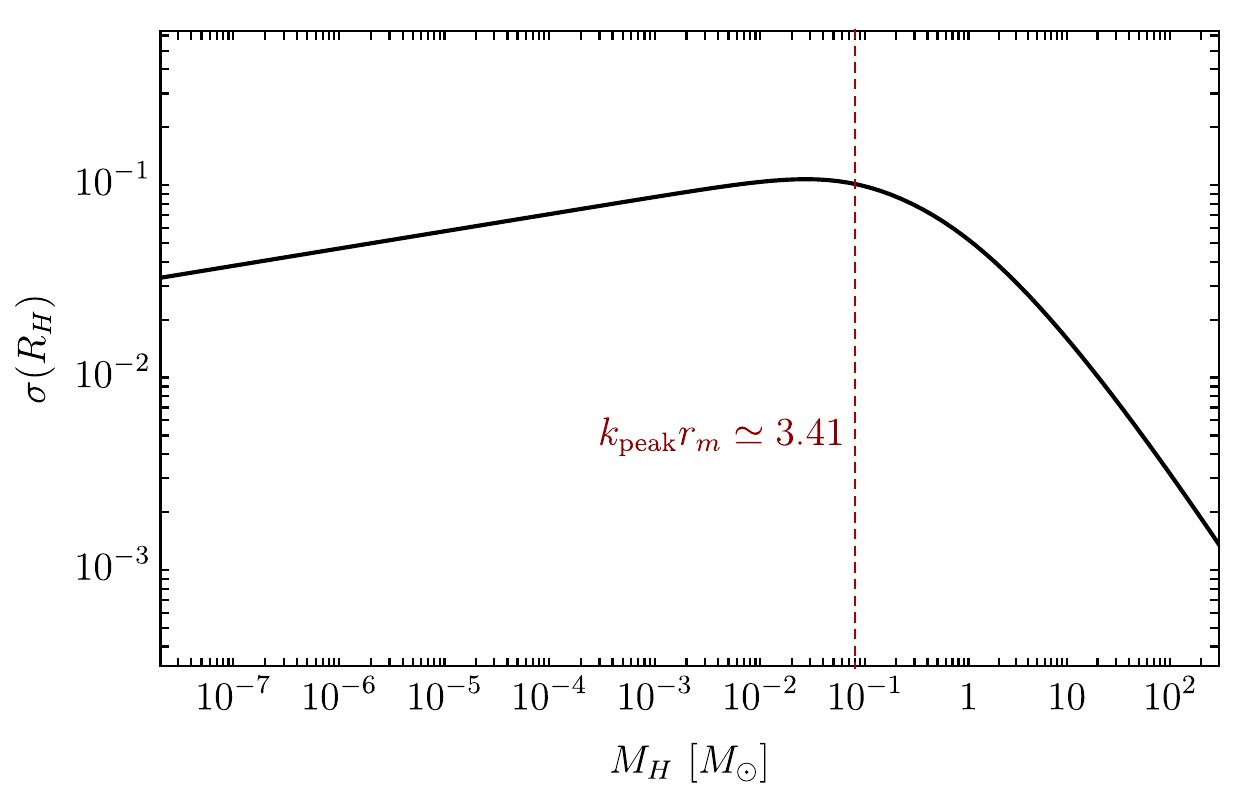}
	\caption{
Variance of the density contrast filed in 
eq.\,(\ref{eq:Sigma}) as a function of 
$M_H$. 
We take $\Phi = 2/3$ (that is, radiation-dominated universe). 
 }
\label{fig:SigmaSqPlot}
\end{figure}
Within this choice, we show our result for the variance of the smoothed density contrast field in fig.\,\ref{fig:SigmaSqPlot} as a function of the horizon mass.
We observe that the variance exhibits a maximum value for modes with a comoving wavenumber at around $k_H^{-1} \approx 3.41/k_{\textrm{peak}}$. This corresponds to the horizon mass $M_H \simeq 0.1\,M_{\odot}$. 
This is expected.
As we have discussed, $3.41/k_{\textrm{peak}}$ is  the
typical scale of a collapsing perturbation in real space. 
In Fourier space,  modes with comoving wavenumber at around the inverse of this value are those that will contribute the most to the collapse.

\begin{figure}[!t]
	\centering
\includegraphics[width=0.495\textwidth]{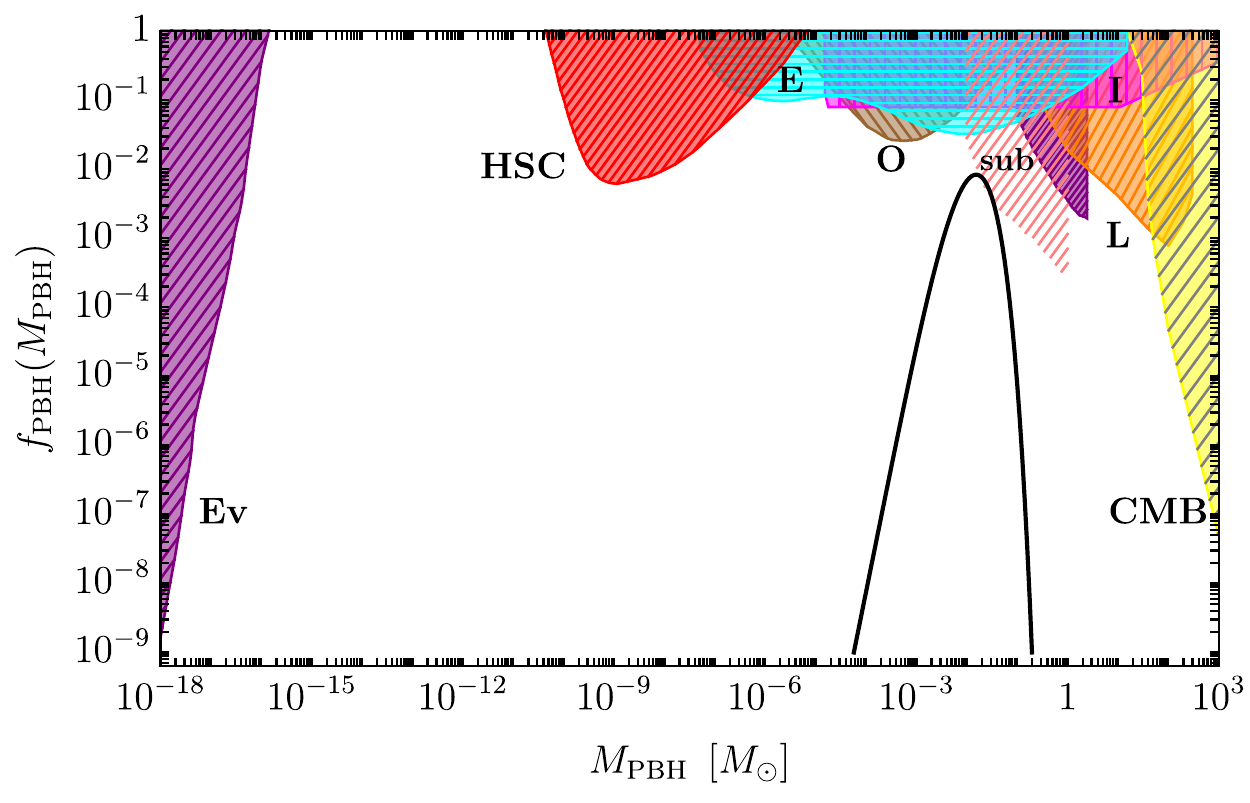}
	\caption{
 Fraction of dark matter in the form of PBHs with mass $M_{\textrm{PBH}}$
(see ref.\,\cite{Green:2020jor} for a review and\,\href{github.com/bradkav/PBHbounds}{\faGithub/bradkav/PBHbounds}  for an updated repository of data). We show the following bound.
\textbf{Ev}aporation constraints (see also \cite{Saha:2021pqf,Laha:2019ssq,Ray:2021mxu}): EDGES\,\cite{Mittal:2021egv}, 
CMB\,\cite{Clark:2016nst}, INTEGRAL\,\cite{Laha:2020ivk,Berteaud:2022tws}, 511 keV\,\cite{DeRocco:2019fjq,Dasgupta:2019cae}, Voyager\,\cite{Boudaud:2018hqb}, 
EGRB\,\cite{Carr:2009jm};
microlensing constraints from the Hyper-Supreme Cam (\textbf{HSC}), ref.\,\cite{Niikura:2017zjd}; 
microlensing constraints from \textbf{E}ROS, ref.\,\cite{EROS-2:2006ryy}; 
microlensing constraints from \textbf{O}GLE, ref.\,\cite{Niikura:2019kqi}; 
\textbf{I}carus microlensing event, ref.\,\cite{Oguri:2017ock}; 
constraints from modification of the \textbf{CMB} spectrum due to accreting PBHs, ref.\,\cite{Serpico:2020ehh};
direct constraints on PBH-PBH mergers with \textbf{L}IGO, refs.\,\cite{LIGOScientific:2019kan,Kavanagh:2018ggo} (see also \cite{Wong:2020yig,Hutsi:2020sol,DeLuca:2021wjr,Franciolini:2021tla,Franciolini:2022tfm}); 
search for GWs from the coalescence of \textbf{sub}-solar mass 
binaries, ref.\,\cite{Nitz:2021vqh}. 
 }
\label{fig:PBHAbu}
\end{figure}

After the change of variables  from $\delta_{\rm L}$ to $M_{\rm PBH}$ (cf. ref.\,\cite{Byrnes:2018clq}),
 we arrive at the final formula
\begin{align}\label{eq:massfunctionintegral}
&f_{\rm PBH}(M_{\rm PBH}) = \frac{1}{\Omega_{\rm CDM}}\int_{M_H^{\rm min}}^{\infty}
\left(
\frac{M_{\rm eq}}{M_H}
\right)^{1/2} \times \nn\\&
\frac{
e^{-\frac{8}{9\sigma^2(M_H)}\left[
1 - 
\sqrt{\Lambda}\,
\right]
^2}
}{
\sqrt{2\pi}\sigma(M_H)
\Lambda^{1/2}}\left(
\frac{M_{\rm PBH}}{\gamma M_H}
\right)\left(
\frac{M_{\rm PBH}}{\mathcal{K}M_H}
\right)^{1/\gamma}d\log M_H
\,,
\end{align}
where we defined
\begin{equation}
    \Lambda \equiv 1 - \frac{\delta_{\textrm{th}}}{\Phi} - \frac{1}{\Phi}\left(\frac{M_{\rm PBH}}{\mathcal{K}M_H}\right)^{1/\gamma}\,.
\end{equation}
In eq.\,(\ref{eq:massfunctionintegral}), the  lower limit of integration follows from the condition $\Lambda > 0$.
We integrate numerically eq.\,(\ref{eq:massfunctionintegral}). Our result is shown in fig.\,\ref{fig:PBHAbu}.
The parameters of the models are tuned in such a way to saturate the existing constraints (cf. caption of fig.\,\ref{fig:PBHAbu} for details). 
As expected, the  mass distribution of PBHs in our model peaks in the sub-solar mass range. In this mass range, it is possible that a fraction of a percent of the dark matter is constituted by PBHs. 
In the case of our model the most stringent constraint comes from the search for GWs from the merger of sub-solar mass binaries in the first half of the Advanced LIGO and Virgo’s third observing run\,\cite{Nitz:2021vqh}. 
The sub-solar mass range is particularly intriguing since standard stellar evolution processes do not account for the formation of sub-solar mass black holes.
The detection of even a single sub-solar mass black hole would provide crucial evidence supporting the existence of PBHs.

As a final remark, we checked that including the effect of the QCD phase transition following refs.\,\cite{Franciolini:2022tfm,Musco:2023dak}
has a very small impact on our numerical results. 
As anticipated, this is because the PBH mass distribution peaks below the solar mass range.

\subsection{Scalar-induced gravitational waves}\label{sec:SIWG}

\begin{figure}[!t]
	\centering
\includegraphics[width=0.495\textwidth]{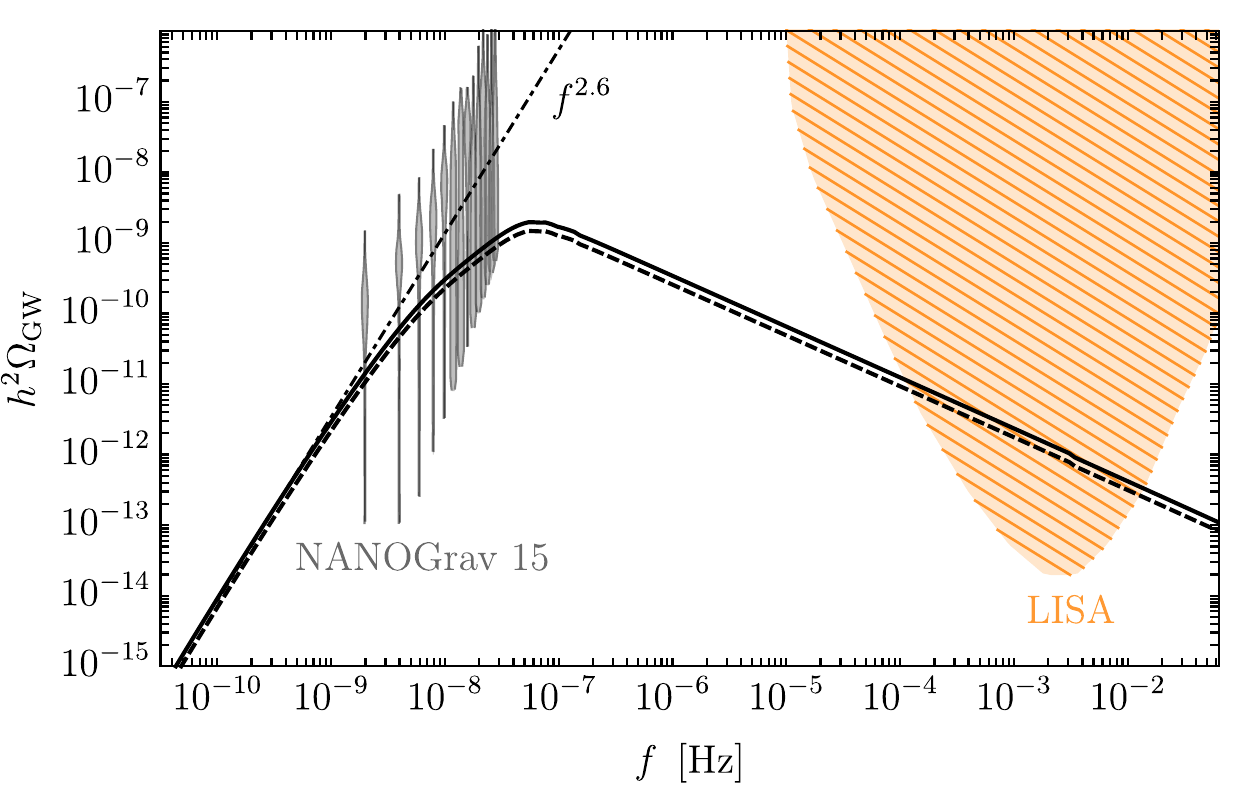}
	\caption{	
Fraction of the energy density in GWs relative to the critical energy density of the universe as a
function of the frequency. 
We show the power-law integrated sensitivity curve of the future space-based GW interferometer LISA (region hatched in orange, cf. 
ref.\,\cite{LISA:2017pwj}). 
The gray violins correspond to
the posteriors of an Hellings-Downs-correlated free spectral reconstruction of the NANOGrav\,15 signal\,\cite{NANOGrav:2023gor,NANOGrav:2023hvm}.
We use as a comparison the first 14 bins of the NANOGrav\,15 signal, thus following what the collaboration itself has identified as the signal attributed to a stochastic GW background.
The solid black line corresponds to the scalar-induced GW signal in our model. 
The dashed black line takes into account the presence of primordial non-Gaussianity, cf. section\,\ref{sec:Discu}. 
For illustrative purposes, the dot-dashed line corresponds to a power-law fit of the far tail of the scalar-induced GW signal.
 }
\label{fig:GWBPlot}
\end{figure}

\begin{figure}[!t]
	\centering
\includegraphics[width=0.495\textwidth]{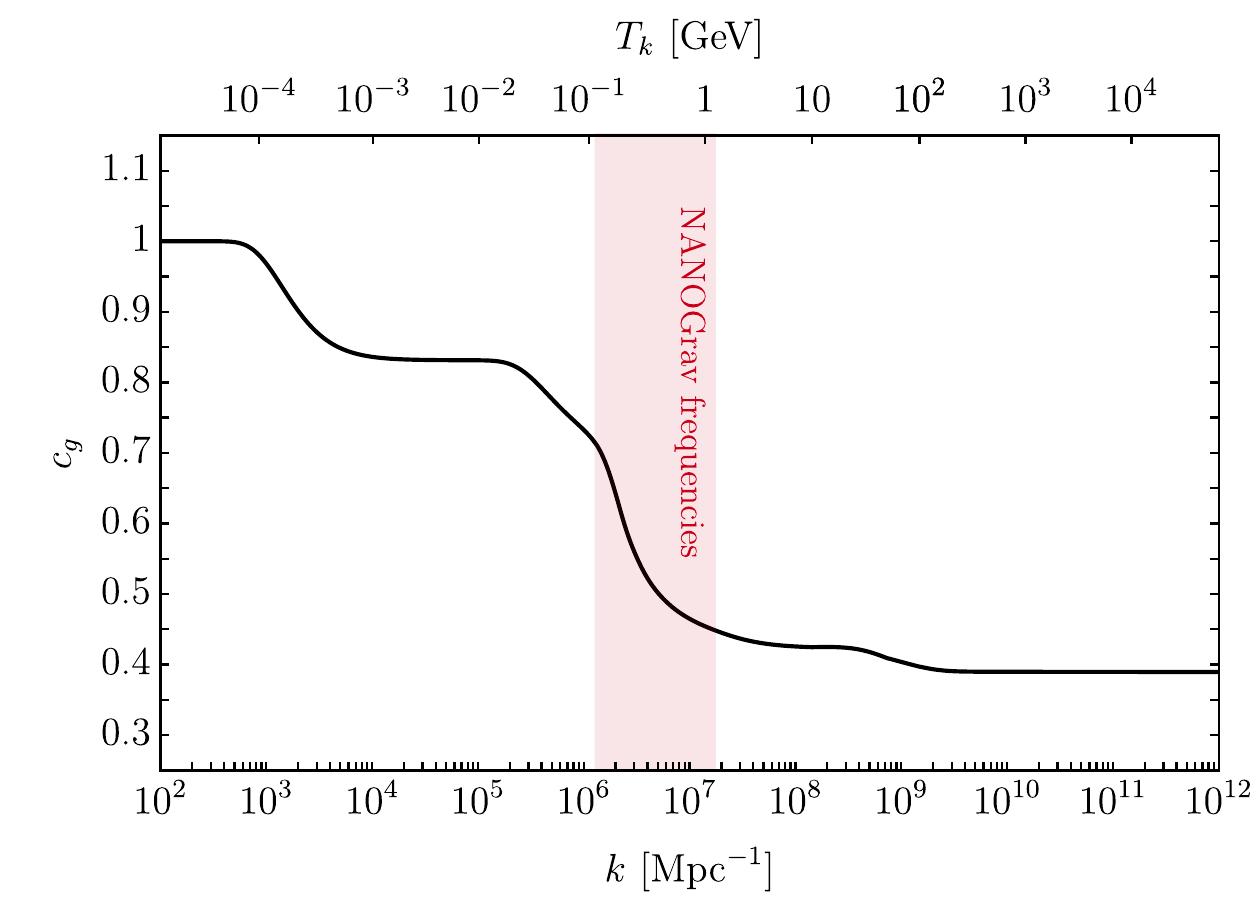}
	\caption{	
Evolution of the parameter $c_g$ defined in eq.\,(\ref{eq:cgPar}) as function of $k$ (bottom $x$-axis) and temperature $T_k$  (top $x$-axis), cf. eq.\,(\ref{eq:Temperaturek}). 
The vertical band corresponds to the values of $k$ that are relevant for the NANOGrav\,15 signal.
 }
\label{fig:cgPlot}
\end{figure}

Given the power spectrum of scalar perturbations computed by means of eq.\,(\ref{eq:M-S}) and shown in fig.\,\ref{fig:PowerSpectrum}, we compute
the induced second-order GW spectrum. Our result is shown in fig.\,\ref{fig:GWBPlot} where we plot
the fraction of the energy density of gravitational waves relative to the critical energy density, $\Omega_{\textrm{GW}}$. 
We superimpose the prediction of our model  to the NANOGrav\,15 signal (cf. the caption for details). 
We follow the
computation recently revisited in ref.\,\cite{Espinosa:2018eve}, and we refer to this paper for details (cf. also ref.\,\cite{Franciolini:2022pav}). In what follows, we focus on highlighting the most relevant points for our analysis.
Let us write the amplitude of induced GW spectral density measured today in the form~\cite{Domenech:2021ztg}
\begin{align}
\Omega_{\rm GW}(f) & = \frac{c_g\Omega_r}{36}\int_{0}^{\frac{1}{\sqrt{3}}}dt\int_{\frac{1}{\sqrt{3}}}^{\infty}ds
T(s,t)
P_{\zeta}(ku)
P_{\zeta}(kv)\,,\label{eq:NGGW}\\
T(s,t) & \equiv \left[
\frac{(t^2-1/3)(s^2-1/3)}{t^2 - s^2}
\right]^{2}\left[\mathcal{I}_c(t,s)^2 + \mathcal{I}_s(t,s)^2\right]\,,
\label{eq:Tra}
\end{align}
where $\Omega_r$ is the current energy density of radiation, $u\equiv \sqrt{3}(s+t)/2$, $v\equiv \sqrt{3}(s-t)/2$,
and $\mathcal{I}_c$ and $\mathcal{I}_s$ are two functions that  can  be  computed  analytically (see, for instance, refs.\,\cite{Espinosa:2018eve,Kohri:2018awv}). 
The parameter $c_g$, defined as 
\begin{equation}
c_g	\equiv \frac{g_*(T_k)}{g_{*}^0}
\left[\frac{g_{*S}^0}{g_{*S} (T_k)}\right]^{4/3}\,,\label{eq:cgPar}
\end{equation}  accounts for the change of the effective degrees of freedom of the thermal radiation $g_*$ and $g_{*S}$ in the early universe (assuming Standard Model physics). 
The superscript $^0$ indicates the values of $g_*$ and $g_{*S}$ today while 
$g_*(T_k)$ and $g_{*S} (T_k)$ are evaluated at the time of horizon crossing (in this case horizon re-entry since we are after the end of inflation) of mode $k$. 
Each mode $k$ crosses
the horizon at the temperature $T_k$ given by the relation
\begin{align}
k = 1.5\times 10^7
\left(\frac{g_*}{106.75}\right)^{\frac{1}{2}}
\left(\frac{g_{*,s}}{106.75}\right)^{-\frac{1}{3}}
\left(\frac{T_k}{\textrm{GeV}}\right)
\textrm{Mpc}^{-1}\,.\label{eq:Temperaturek}
\end{align}
As far as the temperature-dependence of $g_*$ and $g_{*,s}$ we refer to ref.\,\cite{Saikawa:2018rcs}. 
Eq.\,(\ref{eq:Temperaturek}) shows that 
the wavenumbers that are relevant for the NANOGrav\,15 signal re-enter the horizon at around the time of the QCD phase transition. 
Deep in the radiation epoch, we have $c_g \approx 0.4$.  The temperature-dependence of $c_g$ is shown in fig.\,\ref{fig:cgPlot}.
The values of $k$ that are relevant for the NANOGrav\,15 signal corresponds to temperature below 1 GeV and the corresponding value of $c_g$ is slightly larger than $0.4$. 
We implement the temperature-dependence of $c_g$ in our numerical computation.  
Notice that, in principle, one should also modify the transfer function in 
eq.\,(\ref{eq:Tra}) since the latter is written assuming pure radiation. 
This has been done in refs.\,\cite{Abe:2020sqb,Ferrante:2023bgz}; the effect is minor, and we do not include it in our analysis. 
For a more detailed discussion regarding the impact of the QCD phase transition on the spectrum of the induced GW signal, we refer the interested reader to refs.\,\cite{Hajkarim:2019nbx,Franciolini:2023wjm}.

As far as the comparison with the NANOGrav\,15 signal in concerned, we will engage in a more critical and thorough discussion of our findings in section\,\ref{sec:Discu}. 
For the time being, we confine our observation to the fact that the part of the scalar-induced GW signal that crosses the frequencies observed by NANOGrav\,15 lies between the peak and the low-frequency tail, cf. fig.\,\ref{fig:GWBPlot}. 
In the low-frequency tail far from the peak, we find that our scalar-induced GW signal is well described by a power-law with spectral index $\sim f^{2.6}$. 
We checked that this result is in agreement with the analytical estimates in the scenario where the power spectrum follows a broken power law\,\cite{Domenech:2021ztg}. 
We notice that such value of the spectral index is quite aligned with the one that can be inferred from NANOGrav\,15 data (cf. discussion below eq.\,(\ref{eq:OmegaPar})) particularly when restricting our attention to the first few bins of the measured signal (where data are more precise). 
As shown in our fig.\,\ref{fig:GWBPlot}, as we approach the peak, the slope of the scalar-induced GWs tends to decrease. However, within this frequency range, the comparison with the data is subject to greater uncertainties, and such a frequency-dependent trend cannot be ruled out at present. 
What appears to be more problematic is the amplitude of the signal, which visually appears to be smaller than what would be required to fit the data. We will discuss this point in section\,\ref{sec:Discu}.

\subsection{Conclusive remarks}\label{sec:comments}

We conclude this section with some remarks concerning our model from the perspective of the effective QFT.
\begin{itemize}
\item[{\it (i)}] The amplitude $A_s$ of the scalar power spectrum at CMB scales gives access to the combination of parameters $a_4 M^2/\MPl^2$. Consequently, by imposing $A_s = 2.12\times 10^{-9}$, we find the value $a_4 M^2/\MPl^2  = O(10^{-10})$ reported in table\,\ref{tab:Parameters}.

This is quite a small number; to interpret it, 
we propose that a criterion for our theory is that pure (dimensionless) quantities should be of order $O(1)$. 
Specifically, we demand that this condition holds true for the dimensionless parameters $a_{2,3,4}$ and $\xi$ that characterize the renormalizable part of our Lagrangian density in eq.\,(\ref{eq:Lagra}).
Under this assumption, the theory in eqs.\,(\ref{eq:Lagra},\,\ref{eq:Pote}) can be considered natural in the Wilsonian sense. 
If we posit $a_4 = O(1)$, we see that the small value suggested by $A_s$ is now interpreted as $M/\MPl = O(10^{-5})$. 
This is interesting because, as we discussed in eq.\,(\ref{eq:String}), the ratio $M/\MPl$ has the dimension of a coupling. 
In other words, the constraint on $A_s$ can be interpreted as indicating that the ultraviolet completion from which our scalar effective QFT arises is a weakly coupled theory. 
\item[{\it (ii)}] 
As far as the parameters $a_{2,3}$ are concerned, we find 
\begin{align}
a_2 \simeq  2.2\,a_4\,,~~~~~~a_3 \simeq  -2.5\,a_4\,.
\end{align}
These values are opposite in sign; this is expected, as they must conspire to yield the stationary inflection point. 
Furthermore, they are both of order $O(1)$. 
\item[{\it (iii)}] 
The Hubble parameter in fig.\,\ref{fig:ClassicalDyn} is of order $H = O(10^{-5})\,M \ll M$.   
The Hubble parameter $H$ provides a measure of the inflationary energy scale, while $M$ represents the mass scale characterizing the states of the sector that have been integrated out to construct our effective theory. 
The condition $H \ll M$ is, therefore, highly favorable, as it signifies that during inflation, there isn't sufficient energy to excite degrees of freedom with mass of order $M$. 
Combined with point{\it (ii)} above, we expect some hierarchy of the type $H\ll M\ll \MPl$.
\item[{\it (iv)}] The coefficients of the higher-dimensional operators included in our analysis are given by (cf. table\,\ref{tab:Parameters}) 
$a_5 \simeq -6\times 10^{-2}\,a_4$, $a_6  \simeq 2\times 10^{-3}\,a_4$. Assuming $a_4 = O(1)$, we have 
$a_{5,6}\ll 1$. This seems to contradict point {\it (i)} above. However, it is important to bear in mind that $a_{5,6}$ represent coefficients of higher-dimensional operators, and as such, they could potentially be generated at the loop level. In the latter scenario, the coefficients $a_{5,6}$ would naturally be accompanied by a suppression factor of order $\kappa g^2$ with $\kappa = 1/(4\pi)^2$.

Finally, we note that the coefficients $a_{n}$ enter with the peculiar pattern of alternating sign $(-1)^n$. 
This information could potentially shed some light on the ultraviolet completion of our theory.
\end{itemize}
Clearly, it is unnecessary to emphasize that the discussion elaborated upon in the preceding points does not intend to claim exhaustiveness or meticulous precision. On the other hand, our effective theory is formulated under simplifying assumptions (in particular, we have assumed that the sector we integrated out is characterized by a single coupling and a single mass scale). Nonetheless, it is intriguing to observe how inflationary observables manage to provide some insights into the properties of the underlying theory.

\section{Discussion and outlook}\label{sec:Discu}

In this work, we have formulated a single-field model of inflation that demonstrates compatibility, at CMB scales, with the BK18 dataset as well as the Planck observations. 
Conceptually, the main ingredients characterizing our model are:
\begin{itemize}
    \item[$\ast$]  the presence of a non-minimal coupling to gravity;
    \item[$\ast$] the inclusion of effective operators up to (at least) dimension 6 with alternating signs;
    \item[$\ast$] the presence of an approximate stationary inflection point in the Einstein frame potential.
\end{itemize}  
The model, through the generation of second-order GWs, has the potential to actively participate in shaping the recently detected nHz GW signal by the NANOGrav collaboration. 
In this section, we delve more extensively into this last statement. 

In our model, the scalar-induced GW signal shown in fig.\,\ref{fig:GWBPlot} 
has an amplitude which is actually slightly smaller than the one needed to give a perfect fit of NANOGrav\,15 data. 
Even without the need for running numerical analysis, it becomes quite evident from fig.\,\ref{fig:GWBPlot} that the scalar-induced GW signal is marginally compatible with the NANOGrav\,15 data. 
At first glance, this conclusion appears rather robust, as attempting to enhance the GW signal would inevitably lead to an overproduction of sub-solar PBH mergers in tension with osbervations, cf. fig.\,\ref{fig:PBHAbu}.  
This conclusion is in agreement with the  analysis of the scalar-induced GW signal 
investigated in ref.\,\cite{NANOGrav:2023hvm} (see also refs.\,\cite{Dandoy:2023jot,Franciolini:2023pbf}). 

However, the message we intend to convey through our analysis should by no means be regarded as negative.  
First and foremost, as discussed in the introduction, it is entirely plausible to anticipate that the NANOGrav\,15 signal contains a component attributable to SMBHB.  Consequently, it does not appear strictly necessary to pursue a perfect fit of the signal using exclusively a new physics channel. Conversely, attaining a signal that can contribute significantly, albeit not comprehensively, constitutes an inherently intriguing outcome, particularly in the context of a possible multi-component analysis.

Secondly, there exist certain theoretical uncertainties that render the computation of the scalar-induced GW signal not yet entirely under complete control.
The crucial point resides in the fact that, once the model parameters have been chosen to reproduce BK18 constraints and those related to the PBH abundance, the computation of the scalar-induced GW signal transforms into a bona fide prediction. 
This represents both a curse and a blessing. 
A blessing, as it signifies the possibility to falsify the model through correlated observations; a curse, since it necessitates the most reliable calculation of the PBH abundance. 

It is important to have a clear understanding of the significance and limitations of these two statements. Firstly, regarding the potential to falsify models through correlated observations, it should be emphasized that the existence of a  GW signal arising from the merger of two sub-solar mass PBHs is to be regarded as a sufficient yet not necessary condition for the presence of a scalar-induced GW signal. The underlying reason is that the abundance of PBHs exhibits exponential sensitivity to the amplitude of the power spectrum. A marginal reduction in the latter by a small percentage would lead to a decrease in the PBH merger rate to undetectably small levels while leaving the amplitude of the scalar-induced GW signal (which depends only quadratically on the amplitude of the power spectrum) relatively unaffected.

Secondly, concerning the computation of the PBH abundance, there exist a number of theoretical effects that cannot yet be deemed entirely under control. 
Let us comment on some of them; we highlight in {\color{harvardcrimson}{red}} ({\color{azure}{azure}}) the effects that could attenuate (enhance) the amplitude of scalar-induced GWs.

{\color{harvardcrimson}{{\underline{Primordial non-Gaussianity}}}}. We computed the quantity $f_{\rm PBH}(M_{\rm PBH})$ 
in eq.\,(\ref{eq:massfunctionintegral}) assuming Gaussian statistics for the comoving curvature perturbation field $\zeta$. 
However, the presence of local non-Gaussianity 
of primordial origin -- typically parameterized through the parameter $f_{\textrm{NL}}$\footnote{At the quadratic level, $f_{\textrm{NL}}$ is defined by the relation $\zeta = \zeta_{\textrm{G}} + (3/5)f_{\textrm{NL}}\zeta_{\textrm{G}}^2$, where $\zeta_{\textrm{G}}$ follows a Gaussian statistics.} -- is unavoidable since the 
curvature perturbation field is governed by  non-linear dynamics. The true question is whether or not the presence of such primordial non-Gaussianity gives a sizable effect in the computation of the PBH abundance. 
In principle, generalizing eq.\,(\ref{eq:massfunctionintegral}) to include the presence of local non-Gaussianity in the curvature perturbation field is relatively straightforward by means of the formalism developed in ref.\,\cite{Ferrante:2022mui} (see also refs.\,\cite{Young:2022phe,Gow:2022jfb}). 
This analysis has been carried out 
in ref.\,\cite{Franciolini:2023pbf}. 
The findings of this analysis are consistent with the expectations drawn from prior literature\,\cite{Young:2013oia,Atal:2018neu,Taoso:2021uvl,Riccardi:2021rlf}. The presence of primordial non-Gaussianity with $f_{\textrm{NL}} > 0$ (as expected in the case of single-field inflationary models with USR dynamics) tends to facilitate the formation of PBHs, and one is forced to consider a  lower peak value in the scalar power spectrum. Consequently, this results in a smaller amplitude of the scalar-induced GW signal that tends to worsen the comparison with NANOGrav\,15 data. 

In this context, the point we would like to make is that within the framework of realistic models of single-field inflation featuring USR dynamics and aimed at reproducing the NANOGrav\,15 signal, the amount of primordial non-Gaussianity is bound to be quite small (cf. also ref.\,\cite{Atal:2021jyo}). Our back-of-the-envelope computation goes as follows. 
From eq.\,(\ref{eq:PSSR}), assuming constant $H$, we get 
$\epsilon_{\textrm{\tiny{USR}}} \approx \epsilon_{\star}A_s/A_{\textrm{peak}}$ which relates the value of $\epsilon$ at around the USR phase, $\epsilon_{\textrm{\tiny{USR}}}$, to the value of $\epsilon$ at the pivot scale, $\epsilon_{\star}$, and the ratio 
$A_s/A_{\textrm{peak}}$ between the amplitude of the scalar power spectrum at the pivot scale, $A_s$, and the peak amplitude  of the power spectrum attained during the USR 
phase, $A_{\textrm{peak}}$. 
We use $A_s \simeq 2\times 10^{-9}$ while we relate $\epsilon_{\star}$ to the tensor-to-scalar ratio via $r \approx 16\epsilon_{\star}$. We then find
\begin{align}
  \epsilon_{\textrm{\tiny{USR}}} \approx 
  5\times 10^{-10}\left(\frac{r}{0.036}
  \right)
\left(\frac{10^{-2}}{A_{\textrm{peak}}}\right)\,.
\end{align}
After the end of USR, $\epsilon$ starts growing exponentially fast according to 
$\epsilon(N) \sim e^{-2\eta_{0} N}$ 
where $\eta_{0}$ represents the (approximately constant) negative value of $\eta$ after the end of the USR phase (cf. fig.\,\ref{fig:ClassicalDyn}). 
Inflation ends when $\epsilon = O(1)$. 
The previous scaling, therefore, implies that 
it takes 
\begin{align}
\Delta N \approx \frac{1}{2\eta_0}\log(\epsilon_{\textrm{\tiny{USR}}})
\,,\label{eq:forv}
\end{align}
$e$-folds for $\epsilon$ to reach $O(1)$ values starting from  $\epsilon_{\textrm{\tiny{USR}}}$. 
As we discussed in section\,\ref{sec:PS}, in order to have a scalar-induced GW signal at around PTA frequencies, we need to place the peak of the scalar power spectrum at around $20$ $e$-folds  after horizon crossing of the pivot scale $k_{\star}$. This means that, to conclude inflation in a way that solves the cosmological problems we need $40$-ish $e$-folds more of accelerated expansion after the USR phase. Consequently, we rewrite eq.\,(\ref{eq:forv}) as 
\begin{align}
\eta_0 & \approx \frac{1}{2\Delta N}\log(\epsilon_{\textrm{\tiny{USR}}}) \nn\\
& \approx 
\frac{1}{2\Delta N}\log\bigg[
5\times 10^{-10}\left(\frac{r}{0.036}
  \right)
\left(\frac{10^{-2}}{A_{\textrm{peak}}}\right)
\bigg]\,.
\end{align}
For the benchmark values $r=0.036$ and $A_{\textrm{peak}} = 10^{-2}$, 
$\Delta N = 40$  corresponds to $\eta_0 \approx -0.27$. 
Despite the crude approximations, this value is in good agreement with what we found in our numerical analysis ($\eta_0 \simeq -0.2$, cf. fig.\,\ref{fig:ClassicalDyn}). 
The relevance of this estimate is that the value of $\eta_0$ controls the size of the local primordial non-Gaussianity since we have $f_{\textrm{NL}} = -5\eta_0/6$\,\cite{Atal:2018neu,Taoso:2021uvl}; consequently, $\eta_0 \simeq -0.2$ implies $f_{\textrm{NL}} \simeq 0.167$ which is indeed quite a small value. In the language of the parameter $\beta$ used in ref.\,\cite{Franciolini:2023pbf}, we find 
$\beta = -2\eta_0 \simeq 0.4 \ll 3$, where $\beta = 3$ is the benchmark value used in ref.\,\cite{Franciolini:2023pbf} to visualize the impact of non-Gaussianity on the  scalar-induced GW signal in models with an approximate stationary 
inflection point. It should be also noted that $\eta_0$ controls the shape of the power spectrum at large $k$, the latter scaling as $P_{\zeta}(k) \sim k^{2\eta_0}$\,\cite{Riccardi:2021rlf}. 
From this perspective, it is clear that if the parameter $-\eta_0$ is too large, the power spectrum -- after the peak located at around $10^{7}$ Mpc$^{-1}$ as required to match PTA frequencies -- will decay too rapidly, causing inflation to conclude too swiftly.

In conclusion, we expect the impact of primordial non-Gaussianity on the PBH abundance to be rather small in our model. 
Given the previous argument, we expect  the same conclusion to be valid in all single-field inflationary models that generate a sizable scalar-induced GW signal via USR at PTA frequencies. 
In order to confirm this expectation, 
we include the effect of primordial non-Gaussianity using the formalism developed in ref.\,\cite{Ferrante:2022mui}. We take the functional form\,\cite{Atal:2019cdz,Biagetti:2021eep} 
$\zeta = -(6f_{\textrm{NL}}/5)^{-1}
\log[1-(6/5)f_{\textrm{NL}}\zeta_{\textrm{G}}]
$, where $\zeta_{\textrm{G}}$ is the Gaussian component 
of the curvature perturbation field with 
$f_{\textrm{NL}} = 0.167$ as discussed before. 
We re-tune the value of the non-minimal coupling $\xi$ in order to saturate the allowed abundance of PBHs, and we re-compute the predicted signal of scalar-induced GWs. 
Our final result corresponds to the dashed black line in fig.\,\ref{fig:GWBPlot}.
As anticipated, the effect is minor, resulting in a rescaling of the signal computed without the inclusion of primordial non-Gaussianity by a factor of $3/4 = 0.75$. 

Finally, notice that, in principle, the presence of primordial non-Gaussianity could also impact the determination of the threshold for gravitational collapse (that we computed, following ref.\,\cite{Musco:2020jjb}, including only the non-
linear relation between the density contrast and the curvature perturbation field). 
This effect was investigated in refs.\,\cite{Kehagias:2019eil,Escriva:2022pnz} (although in the idealized case of a monochromatic power spectrum). 
For the small values of $f_{\textrm{NL}}$ relevant for this work, the effect seems to be small. 

{\color{harvardcrimson}{{\underline{Peak theory versus threshold statistics}}}}. 
We computed the abundance of PBH assuming Press–Schechter threshold statistics. 
This is manifest in eq.\,(\ref{eq:BetaDef}) in which the mass fraction $\beta$ was computed as
the probability that density contrast is larger than the threshold for PBH formation.  
An alternative approach based on peak theory\,\cite{Bardeen:1985tr} could equally be employed\,\cite{Green:2004wb}. 
 It is known that threshold statistics
gives a smaller PBH abundance if compared with peak theory (see, e.g., ref.\,\cite{DeLuca:2019qsy}). The difference between threshold statistics and peak theory is exacerbated to a greater extent as the power spectrum becomes more peaked\,\cite{Riccardi:2021rlf}.
At equal abundance, therefore, a peak theory-based approach would necessitate a lower peak amplitude of the power spectrum compared to that required by threshold statistics.

{\color{azure}{{\underline{Non-linearities at horizon
crossing}}}}. 
This is potentially the most sensitive point. 
As we have already mentioned in section\,\ref{sec:Tech}, the threshold parameter in eq.\,(\ref{eq:ThresholdParameter}) is typically calculated by linearly extrapolating the value obtained at super-horizon scales to the horizon crossing.  
As discussed in ref.\,\cite{Musco:2020jjb}, non-linear corrections to the above extrapolation tend to double the value of the threshold thus making the production of PBHs more difficult. 
As a result, to achieve the same PBH abundance, one would need to enhance the peak amplitude of the scalar power spectrum and, consequently, increase the signal of scalar-induced GWs.
Very recently, this point has been further stressed in ref.\,\cite{DeLuca:2023tun} (see also ref.\,\cite{Germani:2023ojx} for a recent critical discussion about non-linearly estimating the abundance of PBHs). 
Before reaching quantitative conclusions, however, it would be necessary to modify the computation of the PBH abundance. This is certainly an intriguing point that warrants detailed investigation.

{\color{azure}{{\underline{Ellipsoidal Collapse}}}}. 
This is another effect that could increase the threshold for collapse.
The condition for the formation of PBHs is usually investigated within the framework of an isolated spherically symmetric perturbation (with the universe that approaches a state of perfect homogeneity and isotropy on a larger spatial scale).
As discussed in refs.\,\cite{Kuhnel:2016exn,Akrami:2016vrq}, when considering the non-sphericity of the overdensities, the abundance of the resulting PBHs experiences a notable reduction. In this case, the absence of detailed numerical simulations supporting these findings currently prevents us from drawing more quantitative conclusions. Nevertheless, this also appears to be a promising direction worth exploring.

In conclusion, in this study, we have shown that is possible to interpret the NANOGrav\,15 signal within the context of a single-field model of inflation that is in perfect agreement with large-scale CMB measurements. At this juncture, it appears crucial to maximize the refinement of the PBH abundance calculation, which is currently subject to a series of theoretical uncertainties, in order to determine whether the NANOGrav\,15 signal can indeed be genuinely elucidated through inflationary physics.\\

\begin{acknowledgments}
We thank A.~J.~Iovino for discussions.
The work of A.U. is supported in part by the MIUR under contract 2017FMJFMW (PRIN2017).
\end{acknowledgments}

\bibliography{main}

\end{document}